\documentclass[12pt]{article}
\usepackage{latexsym}
\usepackage{amsmath,amsfonts}
\usepackage{times}
\allowdisplaybreaks[4]
\catcode`\@=11

\newcount\hour
\newcount\minute
\newtoks\amorpm \hour=\time\divide\hour by 60\minute
=\time{\multiply\hour by 60 \global\advance\minute by-\hour}
\edef\standardtime{{\ifnum\hour<12 \global\amorpm={am}%
        \else\global\amorpm={pm}\advance\hour by-12 \fi
        \ifnum\hour=0 \hour=12 \fi
        \number\hour:\ifnum\minute<10
        0\fi\number\minute\the\amorpm}}
\edef\militarytime{\number\hour:\ifnum\minute<10 0\fi\number\minute}

\def\draftlabel#1{{\@bsphack\if@filesw {\let\thepage\relax
   \xdef\@gtempa{\write\@auxout{\string
      \newlabel{#1}{{\@currentlabel}{\thepage}}}}}\@gtempa
   \if@nobreak \ifvmode\nobreak\fi\fi\fi\@esphack}
        \gdef\@eqnlabel{#1}}
\def\@eqnlabel{}
\def\@vacuum{}
\def\marginnote#1{}
\def\draftmarginnote#1{\marginpar{\raggedright\scriptsize\tt#1}}
\def\draft{
        \pagestyle{plain}
        \overfullrule=2pt
        \oddsidemargin -.5truein
        \def\@oddhead{\sl \phantom{\today\quad\militarytime} \hfil
        \smash{\Large\sl DRAFT} \hfil \today\quad\militarytime}
        \let\@evenhead\@oddhead
        \let\label=\draftlabel
        \let\marginnote=\draftmarginnote
        \def\ps@empty{\let\@mkboth\@gobbletwo
        \def\@oddfoot{\hfil \smash{\Large\sl DRAFT} \hfil}
        \let\@evenfoot\@oddhead}
        \def\@eqnnum{(\theequation)\rlap{\kern\marginparsep\tt\@eqnlabel}%
        \global\let\@eqnlabel\@vacuum}  }

\newcommand{\rf}[1]{(\ref{#1})}
\renewcommand{\theequation}{\thesection.\arabic{equation}}
\renewcommand{\thefootnote}{\fnsymbol{footnote}}
\newcommand{\newsection}{   % Numeration of eqs. is automatic
\setcounter{equation}{0}\section}

\def\appendix#1{\addtocounter{section}{1}\setcounter{equation}{0}
\renewcommand{\thesection}{\Alph{section}}
\section*{Appendix \thesection\protect\indent \parbox[t]{11.15cm}{#1}}
\addcontentsline{toc}{section}{Appendix \thesection\ \ \ #1}}

\def\be{\begin{equation}}
\def\ee{\end{equation}}
\def\beq{\begin{eqnarray}}
\def\eeq{\end{eqnarray}}

\def\parline{\,\partial\kern -0.55em /\,\,}

\def\half{{\frac{1}{2}}}

\def\LL{{\cal L}}

\def\NN{{\cal N}}

\def\phik{|\phi\rangle}
\def\phibr{\langle\phi|}

\def\ssf{{\sf s}}
\def\ssfb{\bar{\sf s}}

\def\ck{|c\rangle}
\def\cbk{|\bar{c}\rangle}
\def\bk{|b\rangle}

\def\cbbr{\langle \bar{c}|}

\def\bbr{\langle b |}

\def\Lb{\bar{L}}

\def\alphab{\bar{\alpha}}
\def\zetab{\bar{\zeta}}

\def\cb{\bar{c}}
\def\eb{\bar{e}}

\def\rb{\bar{r}}

\def\xik{|\xi\rangle}

\def\oplussm{{\scriptscriptstyle \oplus}}
\def\ominussm{{\scriptscriptstyle \ominus}}

\def\DoF{{\scriptscriptstyle \rm D.o.F}}

\def\oplussm{{\scriptscriptstyle \oplus}}
\def\ominussm{{\scriptscriptstyle \ominus}}

\def\smponetwo{{\scriptscriptstyle [1,2]}}

\def\Awt{\widetilde{A}}

\def\alpar{\alpha\partial}
\def\albpar{\bar\alpha\partial}

\def\irm{{\rm i}}

\def\tr{{\rm tr}}

\def\Irm{{\rm I}}
\def\IIrm{{\rm II}}
\def\Lrm{{\rm L}}

\def\FP{{\rm FP}}
\def\tot{{\rm tot}}
\def\qu{{\rm qu}}
\def\eff{{\rm eff}}
\def\stand{{\rm stand}}
\def\min{{\rm min}}

\def\ibf{{\bf i}}
\def\iibf{{\bf ii}}
\def\iiibf{{\bf iii}}
\def\ivbf{{\bf iv}}
\def\vbf{{\bf v}}
\def\vibf{{\bf vi}}
\def\nbf{{\bf n}}

\def\I{{\rm I}}
\def\II{{\rm II}}

\def\betabf{{\boldsymbol{\beta}}}

\begin{document}

%\draft

\begin{flushright}
FIAN-TD-2016-5 \hspace{2.8cm}{}~\\
arXiv: 1604.02091 V2 [hep-th] \hspace{0.5cm}{}~\\
\end{flushright}

\vspace{1cm}

\begin{center}

{\Large \bf Long, partial-short, and special conformal fields}

\vspace{2.5cm}

R.R. Metsaev\footnote{ E-mail: metsaev@lpi.ru }

\vspace{1cm}

{\it Department of Theoretical Physics, P.N. Lebedev Physical
Institute,
\\ Leninsky prospect 53,  Moscow 119991, Russia }

\vspace{3.5cm}

{\bf Abstract}

\end{center}

In the framework of metric-like approach,
totally symmetric arbitrary spin bosonic conformal fields propagating in flat
space-time are studied. Depending on the values of conformal dimension, spin, and dimension of space-time, we classify all conformal field as long, partial-short, short, and special conformal fields.  An ordinary-derivative (second-derivative) Lagrangian formulation for
such conformal fields is obtained. The ordinary-derivative Lagrangian formulation is
realized by using double-traceless gauge fields, Stueckelberg fields, and auxiliary fields. Gauge-fixed Lagrangian invariant under global BRST transformations is obtained. The gauge-fixed BRST Lagrangian is used for the computation of partition functions for all conformal fields. Using the result for the partition functions, numbers of propagating D.o.F  for the conformal fields are also found.

\newpage
\renewcommand{\thefootnote}{\arabic{footnote}}
\setcounter{footnote}{0}

%%%%%%%%%%%%%%%%%%%%%%%%%%%%%%%%%%%%%%%%%%%%%%%%%%%%%%%%%%%%%%%%%%%%%
\newsection{\large Introduction}
%%%%%%%%%%%%%%%%%%%%%%%%%%%%%%%%%%%%%%%%%%%%%%%%%%%%%%%%%%%%%%%%%%%%%

A study of arbitrary spin conformal fields was initiated in Ref.\cite{Fradkin:1985am}, where a Lagrangian description of totally symmetric conformal fields in space-time $R^{3,1}$ (Fradkin-Tseytlin fields) was developed. A Lagrangian formulation of totally symmetric conformal fields in space-time $R^{d-1,1}$ for  arbitrary $d$ was developed in Ref.\cite{Segal:2002gd}.
Throughout this paper conformal fields studied in Refs.\cite{Fradkin:1985am,Segal:2002gd} will be referred to as short conformal fields. For the reader's convenience, we recall that, in the framework of AdS/CFT correspondence, the short conformal fields in $R^{d-1,1}$ are dual to non-normalizable modes of massless fields in $AdS_{d+1}$. Namely, for spin-2 and spin-$s$, $s\geq 2$, fields it was demonstrated in the respective Ref.\cite{Liu:1998ty} and Ref.\cite{Metsaev:2009ym} that ultraviolet divergence of an action of bulk AdS field evaluated on a solution of the Dirichlet problem coincides with an action of the short conformal field. Besides the short conformal fields, there are conformal fields which we will refer to as long, partial-short and special conformal fields in this paper (for definition, see below). We note then that a minimal Lagrangian formulation of long, short, partial-short and special conformal fields may be found in Ref.\cite{Vasiliev:2009ck}.%
\footnote{ In Ref.\cite{Vasiliev:2009ck}, conformal fields associated with arbitrary  Young tableaux have also been studied. A study of mixed-symmetry conformal fields associated with particular rectangular Young tableaux may be found in Ref.\cite{Marnelius:2009uw}.}

In this paper, we study the long, partial-short and special conformal fields. Such conformal fields are also interesting, among other things, in the context of AdS/CFT correspondence. This is to say that, for arbitrary spin fields, it was demonstrated explicitly in Ref.\cite{Metsaev:2015rda},  that the long conformal field in $R^{d-1,1}$ is dual to non-normalizable modes of massive field in $AdS_{d+1}$ having some discrete value of mass parameter.
Our speculation on string theory interpretation of a conjectural model that involves long higher-spin conformal fields and short low-spin conformal fields may be found in the conclusions of this paper. Before we formulate our main aim in this paper let us discuss a terminology we use throughout this paper.

Consider a free totally symmetric conformal bosonic field propagating in $R^{d-1,1}$. If a Lagrangian of the conformal field is built in terms of one traceless totally symmetric rank-$s$  tensor field of the Lorentz algebra $so(d-1,1)$ then the conformal field will be referred to as spin-$s$ conformal field, while the Lagrangian will be referred to as minimal Lagrangian. If the minimal Lagrangian of free conformal field involves $2\kappa$ derivatives, where $\kappa\geq 1$ is arbitrary integer, then, as well known, a conformal dimension of the conformal field
is given by the expression
\be \label{24022016-01-man}
\Delta = \frac{d}{2} - \kappa\,.
\ee

The use of the labels $\kappa$, $s$, $d$, and our ordinary-derivative approach allows us to classify all conformal fields propagating in $R^{d-1,1}$.
Namely, depending on values of the arbitrary integer $\kappa\geq 1$ and the arbitrary integer $s\geq 1$, conformal fields in $R^{d-1,1}$ with arbitrary $d\geq 3$ and $\Delta$ as in \rf{24022016-01-man} will be referred to as
long, short, partial-short, and special conformal fields. Result of our classification of conformal fields is summarized in the Table (see next page).

For the reader's convenience, we now recall the references devoted to the study of the minimal Lagrangian formulation of conformal fields given in the Table.

\noindent \ibf) For $d=4$ and $d\geq 4$, the minimal Lagrangian of the totally symmetric arbitrary spin short conformal field in the Table was obtained in the respective Ref.\cite{Fradkin:1985am} and Ref.\cite{Segal:2002gd} (see also Ref.\cite{Metsaev:2009ym}).

\noindent \iibf) For $d=4$ and $d\geq4$, the minimal Lagrangian of the totally symmetric arbitrary spin conformal fields with $\kappa=1$ in the Table was obtained in the respective Ref.\cite{Iorio:1996ad} and Ref.\cite{Erdmenger:1997wy}.

\noindent \iiibf) Minimal Lagrangian for all totally symmetric conformal fields given in the Table can be found in Ref.\cite{Vasiliev:2009ck}.

\newpage

\begin{center}
\noindent {\bf Table}. {\sf \small Classification of conformal fields in $R^{d-1,1}$. The integer $s$ indicates spin of \\ conformal field, while the integer $\kappa$ is related to conformal dimension as $\Delta = \frac{d}{2}-\kappa$. }

\bigskip
\begin{tabular}{|c|l|l|}
\hline
&&
\\[-3pt]
{\bf Type of field}  & \hspace{1.5cm} {\bf Values of} $\kappa$  & {\bf Values of} $s$ and $d$
\\[5pt]
\hline
&& \\[-3pt]
long  & $s + \frac{d-4}{2} + N$, \ $N=1,2,\ldots,\infty$  &   \ even $d\geq 4$
\\[3pt]
\cline{1-2}
&& \\[-5pt]
short  & $s + \frac{d-4}{2}$     &  \ $s\geq 1$
\\[3pt]
\hline
&& \\[-3pt]
type II part-short  & $1,2,\ldots, s-1 $  &   \ $d=4$,  \ $s\geq 2 $
\\[3pt]
\hline
&& \\[-3pt]
special  & $ 1,2, \ldots,\frac{d-4}{2} $  &   \ even $d\geq 6$,
\\[3pt]
\cline{1-2}
&& \\[-3pt]
type II part-short  & $\frac{d-2}{2}, \frac{d}{2},\ldots, s $  &    \ $s \geq \frac{d-2}{2} \geq 2$
\\[3pt]
\hline
&& \\[-3pt]
type I part-short  & $ s+1,s+2,\ldots, s + \frac{d-6}{2} $  &   \  even $d \geq 8$
\\[4pt]
& &   \  $s \geq \frac{d-2}{2} \geq 3$
\\[3pt]
\hline
&& \\[-8pt]
special  & $ 1,2, \ldots,s $  &   \ even $d\geq 6$,
\\[3pt]
&&  \ $1 \leq s \leq \frac{d-4}{2}$ \\[5pt]
\hline
&& \\[-8pt]
secondary long &   $s+1, s+2,\ldots,\frac{d-4}{2} $ &  \ even $d \geq 8 $
\\[3pt]
&&    \ $1 \leq s \leq \frac{d-6}{2}$
\\[3pt]
\hline
&& \\[-3pt]
type I part-short  & $  \frac{d-2}{2}, \frac{d}{2}, \ldots,s + \frac{d-6}{2} $  &   \ even  $ d \geq 8$
\\[3pt]
&&  \ $2 \leq s \leq \frac{d-4}{2}$
\\[3pt]
\hline
&& \\[-3pt]
long  & $s +  N$, \ $N=0,1,2,\ldots,\infty$  &   \ $d=3$,  \ $s\geq 1$
\\[3pt]
\hline
&& \\[-3pt]
special  & $1,2,\ldots, s-1$  &   \ $d=3$,  \ $s\geq 2$
\\[3pt]
\hline
&& \\[-3pt]
long  & $s + \frac{d-5}{2} + N$, \ $N=1,2,\ldots,\infty$  &   \ odd $d\geq 5$,  \ $s\geq 1 $
\\[3pt]
\hline
&& \\[-3pt]
special & $1,2,\ldots,s $  &   \ odd $d\geq 5$,  \ $s\geq 1$
\\[3pt]
\hline
&& \\[-3pt]
secondary long  & $s+1,s+2, \ldots, s+ \frac{d-5}{2}$  &  \ odd $d\geq 7$, \ $s\geq 1$
\\[3pt]
\hline
\end{tabular}
\end{center}

\bigskip

With the exception of the particular case $\kappa=1$, the minimal Lagrangian of the totally symmetric conformal fields involves higher-derivatives. Also we note that, with the exception of the short and partial-short conformal fields, the minimal Lagrangian turns out to be gauge variant.%
\footnote{ In this paper, if an action of a conformal field is invariant under gauge transformations, then the respective Lagrangian is referred to as gauge invariant Lagrangian, while, if an action of a conformal field has no gauge symmetries then the respective Lagrangian is referred to as gauge variant Lagrangian.}

Our main aim in this paper is to construct the second-derivative (ordinary-derivative) gauge invariant Lagrangian for all conformal fields given in the Table. We note that the ordinary-derivative description of the short conformal fields was obtained in Refs.\cite{Metsaev:2007fq,Metsaev:2007rw}. In other words, in this paper we extend approach in Refs.\cite{Metsaev:2007fq,Metsaev:2007rw} to the cases of the long, partial-short and special conformal fields. Our Lagrangian formulation of conformal fields has the following two attractive features.

\noindent \ibf) For spin-1, spin-2, and spin-$s$, $s>2$, conformal fields, two-derivative
contributions to our ordinary-derivative Lagrangian take the form of the standard
Maxwell, Einstein-Hilbert, and Fronsdal kinetic terms.

\noindent \iibf) In our approach, all vector and tensor fields are supplemented by appropriate gauge transformations which do not involve higher than first order terms in derivatives. Also we note that the one-derivative contributions to the gauge transformations of all vector and tensor fields take the form of standard gradient gauge transformations.

This paper  is organized as follows.

In Sec. \ref{sec-02},  we summarize conventions and notation we use in this paper.

In Sec. \ref{sec-03}, we briefly review a minimal Lagrangian formulation of spin-$s$ conformal field in $R^{d-1,1}$. We present the minimal Lagrangian for arbitrary values of  $\kappa$, $s$, and $d$ and then we discuss the Lagrangian for some particular values of $\kappa$, $s$, and $d$. Also we discuss how the minimal Lagrangian can be obtained in the framework of AdS/CFT correspondence.

Section \ref{sec-04} is devoted to an ordinary-derivative Lagrangian formulation
of conformal fields. First, we discuss a field content entering our approach. Second, we present our
gauge invariant Lagrangian for all conformal fields in the Table.

In Sec. \ref{sec-05}, we describe gauge symmetries of our ordinary-derivative Lagrangian. We start with a discussion of gauge transformation parameters entering our approach and then we present gauge transformations of conformal fields. In our ordinary-derivative approach, only symmetries of the Lorentz algebra $so(d-1,1)$ are realized manifestly. Therefore, in Sec. \ref{sec-06}, in order to complete our ordinary-derivative formulation, we discuss a realization of the conformal algebra symmetries on a space of the gauge fields entering our approach.

In Sec. \ref{sec-07}, using the Faddeev-Popov procedure, we obtain various representations for ordinary-derivative gauge-fixed BRST Lagrangian. Excluding auxiliary gauge fields and auxiliary Faddeev-Popov fields we obtain a higher-derivative BRST Lagrangian and use such Lagrangian for a computation of partition functions for all conformal fields.

In Sec. \ref{sec-08}, we discuss directions for future research.

%%%%%%%%%%%%%%%%%%%%%%%%%%%%%%%%%%%%%%%%%%%%%%%%%%%%%%%%%%%%%%%%%%%%%%%%%
\newsection{\large Preliminaries}\label{sec-02}
%%%%%%%%%%%%%%%%%%%%%%%%%%%%%%%%%%%%%%%%%%%%%%%%%%%%%%%%%%%%%%%%%%%%%%%%%

%%%%%%%%%%%%%%%%%%%%%%%%%%%%%%%%%%%%%%%%%%%%%%%%%%%%%%%%%%%%%%%%%%%%%%%%%
\subsection{Notation and conventions}
%%%%%%%%%%%%%%%%%%%%%%%%%%%%%%%%%%%%%%%%%%%%%%%%%%%%%%%%%%%%%%%%%%%%%%%%%

Our notation and conventions are as follows. Coordinates of the space-time $R^{d-1,1}$ are denoted by $x^a$, while derivatives with respect to $x^a$ are denoted by $\partial_a$,  $\partial_a \equiv \partial /
\partial x^a$. We use vector indices $a,b,c,e$ of the Lorentz algebra $so(d-1,1)$ which take
the following values $a,b,c,e=0,1,\ldots ,d-1$. Our flat
metric tensor $\eta^{ab}$ is mostly positive. In scalar products, to simplify our expressions we drop
the metric tensor $\eta_{ab}$. In other words, we use the convention $X^aY^a \equiv \eta_{ab}X^a Y^b$.

Throughout this paper a set of creation operators $\alpha^a$, $\alpha^z$, $\zeta$,
$\alpha^\oplussm$, $\alpha^\ominussm$ and the respective set of
annihilation operators $\alphab^a$, $\alphab^z$, $\zetab$,
$\alphab^\ominussm$, $\alphab^\oplussm$ are referred
to as oscillators.
We adopt the following conventions for commutation relations, the vacuum, and hermitian conjugation rules
\beq
&&\hspace{-1.7cm}  [\alphab^a,\alpha^b]=\eta^{ab}\,, \qquad
[\alphab^z,\alpha^z]=1\,, \qquad [\zetab,\zeta]=1\,,
\qquad [\alphab^\oplussm,\, \alpha^\ominussm ]=1\,, \qquad \ [\alphab^\ominussm,\,
\alpha^\oplussm]=1\,,
\\
&& \hspace{-1.7cm} \alphab^a |0\rangle = 0\,,\hspace{1.5cm} \alphab^z |0\rangle = 0\,,\qquad\quad  \zetab|0\rangle =
0\,,\hspace{1cm}   \alphab^\oplussm |0\rangle = 0\,,\hspace{1.4cm}
\alphab^\ominussm |0\rangle = 0\,,
\\
&& \hspace{-1.7cm} \alpha^{a\dagger} =  \alphab^a\,, \hspace{1.6cm}  \alpha^{z\dagger} =
\alphab^z\,, \hspace{1.5cm} \zeta^\dagger = \zetab\,, \hspace{1.3cm} \alpha^{\oplussm\dagger} = \alphab^\oplussm\,,\hspace{1.5cm}
\alpha^{\ominussm \dagger} = \alphab^\ominus \,.
\eeq
The oscillators $\alpha^a$, $\alphab^a$  transform in the vector
representation of the Lorentz algebra $so(d-1,1)$, while
the oscillators $\alpha^z$, $\alphab^z$, $\zeta$, $\zetab$,
$\alpha^\oplussm$, $\alphab^\ominussm$, $\alpha^\ominussm$,
$\alphab^\oplussm$ transform in the scalar representation of the Lorentz algebra.
A hermitian conjugation rule for the derivatives is given by
$\partial^{a\dagger} = - \partial^a$. We use the following shortcuts for operators constructed
out of the oscillators, the derivatives $\partial^a$ and the coordinates $x^a$:
\beq
\label{29022016-06-man} && \hspace{-0.5cm}
\Box
\equiv
\partial^a\partial^a\,,
\qquad\quad  \ \
x\partial
\equiv
x^a \partial^a\,,\qquad\ \
x^2\equiv x^a x^a\,,
\\
\label{29022016-07-man} &&  \hspace{-0.5cm}
\alpha\partial
\equiv
\alpha^a\partial^a\,,
\qquad\quad
\bar\alpha\partial
\equiv \alphab^a\partial^a\,,
\qquad \ \
\alpha^2 \equiv \alpha^a\alpha^a\,,
\qquad\quad   \alphab^2 \equiv
\alphab^a \alphab^a\,,\qquad
\\
\label{29022016-08-man} &&  \hspace{-0.5cm} N_\alpha
\equiv
\alpha^a \alphab^a \,,\qquad \ \ \ \ N_z \equiv \alpha^z
\alphab^z \,,\qquad \ \ N_\zeta \equiv \zeta
\bar\zeta \,,
\\
\label{29022016-09-man} &&  \hspace{-0.5cm}
N_{\alpha^\oplussm}
\equiv
\alpha^\oplussm \alphab^\ominussm\,, \qquad \ \
N_{\alpha^\ominussm}
\equiv \alpha^\ominussm \alphab^\oplussm\,,
\\
\label{29022016-10-man} &&  \hspace{-0.5cm} \Awt^a
\equiv
\alpha^a
- \alpha^2
\frac{1}{2N_\alpha + d-2}\bar\alpha^a \,, \hspace{1.7cm}   \Pi^\smponetwo
\equiv 1 - \alpha^2 \frac{1}{2(2N_\alpha +d)}\bar\alpha^2\,,
\eeq

\beq
\label{28022016-21-man} && r_\zeta = \left(\frac{(s+\frac{d-4}{2} -N_\zeta)(\kappa - s-\frac{d-4}{2}
+ N_\zeta)(\kappa + 1 + N_\zeta)}{2(s+\frac{d-4}{2}-N_\zeta - N_z)(\kappa
+N_\zeta -N_z) (\kappa + 1 + N_\zeta - N_z)}\right)^{1/2}\,,
\\
\label{28022016-22-man} && r_z = \left(\frac{(s+\frac{d-4}{2} -N_z)(\kappa + s + \frac{d-4}{2} -
N_z)(\kappa - 1 - N_z)}{2(s+\frac{d-4}{2}-N_\zeta-N_z)(\kappa + N_\zeta -
N_z) (\kappa -1  + N_\zeta - N_z)}\right)^{1/2}\,.
\eeq

Throughout this paper we adopt the following conventions and notation:
\beq
\label{29022016-05a1-man} && \lambda \in [p]_2 \qquad \Longleftrightarrow \quad \lambda
=-p,-p+2,\ldots, p-2,p\,,
\\
\label{29022016-05a2-man} && \lambda \in [p,q]_1 \quad \Longleftrightarrow \quad \lambda
=p,p+1,\ldots, q-1,q\,,
\\
\label{29022016-05a3-man} && \lambda \in [p,q]_2 \quad \Longleftrightarrow \quad \lambda
=p,p+2,\ldots,  q-2,q\,.
\eeq

%%%%%%%%%%%%%%%%%%%%%%%%%%%%%%%%%%%%%%%%%%%%%%%%%%%%%%%%%%%%%%%%%%%%%%%%%
\subsection{Global conformal symmetries }
%%%%%%%%%%%%%%%%%%%%%%%%%%%%%%%%%%%%%%%%%%%%%%%%%%%%%%%%%%%%%%%%%%%%%%%%%

In the space-time $R^{d-1,1}$, the $so(d,2)$ algebra  is realized as algebra of conformal symmetries.  In a basis of the Lorentz algebra $so(d-1,1)$, the generators of the $so(d,2)$ algebra  are decomposed into
the translation generators $P^a$, the dilatation generator $D$, the conformal
boost generators $K^a$, and the generators of Lorentz algebra $so(d-1,1)$ denoted by  $J^{ab}$.
We use the following commutators of the $so(d,2)$ algebra:
\beq
&& {}[D,P^a]=-P^a\,, \hspace{2.3cm}  {}[P^a,J^{bc}]=\eta^{ab}P^c
-\eta^{ac}P^b \,,
\nonumber\\
&& [D,K^a]=K^a\,, \hspace{2.5cm} [K^a,J^{bc}]=\eta^{ab}K^c -
\eta^{ac}K^b\,,
\nonumber\\
\label{29022016-04-man} && [P^a,K^b]=\eta^{ab}D-J^{ab}\,,  \hspace{1cm} [J^{ab},J^{ce}]=\eta^{bc}J^{ae}+3\hbox{ terms} \,.
\eeq

Consider conformal fields propagating in $R^{d-1,1}$. Let us collect all scalar, vector and tensor fields required for a Lagrangian description of the conformal fields into a ket-vector $\phik$. If a Lagrangian is invariant with
respect to conformal algebra transformation (invariance of a Lagrangian is assumed
to be up to total derivatives)
\be \label{02032016-01-man}
\delta_G \phik  = G \phik
\,,
\ee
then we can present a realization of the conformal algebra generators $G$ in
terms of differential operators acting on $\phik$ in the following way
\beq
\label{02032016-02-man} && P^a
=
\partial^a \,,
\\
\label{02032016-03-man} && J^{ab}
= x^a\partial^b
-  x^b\partial^a + M^{ab}\,,
\\
\label{02032016-04-man} && D
= x\partial
+ \Delta\,,
\\
\label{02032016-05-man} && K^a
=
-\frac{1}{2}x^2\partial^a
+ x^a D
+ M^{ab}x^b + R^a\,.
\eeq
In relations \rf{02032016-03-man}-\rf{02032016-05-man}, $\Delta$ stands for a operator
of conformal dimension, while $M^{ab}$ stands for a spin operator of the Lorentz
algebra $so(d-1,1)$. The operator $M^{ab}$ is acting on spin degrees of freedom collected into the ket-vector $\phik$ and satisfies the following commutation relations:
\be \label{02032016-07-man}
[M^{ab},M^{ce}]
= \eta^{bc}M^{ae}
+ 3\hbox{ terms} \,, \qquad M^{ab}  =  - M^{ba}\,.
\ee
An operator $R^a$ appearing in \rf{02032016-05-man} does not depend on the space-time coordinates $x^a$. In general, this operator depends on the derivatives $\partial^a$.%
\footnote{In the framework of gauge invariant approach to conformal currents and shadow fields
developed in Refs.\cite{Metsaev:2008fs}, the operator
$R^a$ is independent of derivatives $\partial^a$.}
In the framework of minimal Lagrangian formulation of conformal
fields, the operator $R^a$ is equal to zero, while, in the framework of our
ordinary-derivative approach, the operator
$R^a$ turns out to be non-trivial. From relations \rf{02032016-02-man}-\rf{02032016-05-man}, we see that all that is required for the complete description of conformal symmetries is to find a realization of the operators $\Delta$, $M^{ab}$, and $R^a$ on a space of the ket-vector $\phik$.

%%%%%%%%%%%%%%%%%%%%%%%%%%%%%%%%%%%%%%%%%%%%%%%%%%%%%%%%%%%%%%%%%%%%%%
%%%%%%%%%%%%%%%%%%%%%%%%%%%%%%%%%%%%%%%%%%%%%%%%%%%%%%%%%%%%%%%%%%%%%%
\newsection{\large Review of minimal Lagrangian formulation of conformal fields}\label{sec-03}
%%%%%%%%%%%%%%%%%%%%%%%%%%%%%%%%%%%%%%%%%%%%%%%%%%%%%%%%%%%%%%%%%%%%%%
%%%%%%%%%%%%%%%%%%%%%%%%%%%%%%%%%%%%%%%%%%%%%%%%%%%%%%%%%%%%%%%%%%%%%%

To discuss the minimal Lagrangian formulation of a conformal field with arbitrary integer spin $s\geq 1$ and arbitrary integer $\kappa\geq 1$ we use a field $\phi^{a_1\ldots a_s} $ which is totally symmetric traceless rank-$s$ tensor field of the Lorentz algebra $so(d-1,1)$,
\be \label{24022016-04-man}
\phi^{aaa_3\ldots a_s} = 0\,.
\ee
Conformal dimension of the field $\phi^{a_1\ldots a_s}$ is given in \rf{24022016-01-man}. The minimal Lagrangian found in Ref.\cite{Vasiliev:2009ck} can be presented as
\beq
\label{24022016-05-man} \LL  & = &  \half  \sum_{n=0}^N
\frac{2^n  (\kappa +1-n)_n}{  n! (s-n)!(\kappa+s + \frac{d-2}{2}-n)_n }  (\partial^n\phi)^{a_1\ldots a_{s-n}} \Box^{\kappa-n}(\partial^n\phi)^{a_1\ldots a_{s-n}}\,,
\\
\label{24022016-06-man} && (\partial^n\phi)^{a_1 \ldots a_{s-n}} \equiv \partial^{b_1}\ldots \partial^{b_n} \phi^{b_1\ldots b_n a_1\ldots a_{s-n}}\,,
\\
\label{24022016-07-man} && N \equiv \min(s,\kappa)\,,
\\
\label{24022016-08-man} && (p)_q \equiv \frac{\Gamma(p+q)}{\Gamma(p)}\,.
\eeq
We recall that $(p)_q$ defined in \rf{24022016-08-man} is the Pochhammer symbol. From \rf{24022016-05-man} we see that the minimal Lagrangian involves $2\kappa$ derivatives.
For the reader's convenience, we note that the leading terms entering minimal Lagrangian \rf{24022016-05-man} are given by
\beq
\label{24022016-10-man} \LL &\!\!\! = & \!\!\! \frac{1}{2 s!} \phi^{a_1 \ldots a_s} \Box^\kappa \phi^{a_1 \ldots a_s}
\nonumber\\
&\!\!\! + & \!\!\!\frac{\kappa }{(s-1)!(\kappa + s + \frac{d-4}{2})} \partial^b\phi^{ba_1 \ldots a_{s-1}} \Box^{\kappa-1} \partial^c\phi^{ca_1 \ldots a_{s-1}}
\nonumber\\
&\!\!\! + & \!\!\!\frac{\kappa(\kappa-1) }{(s-2)!(\kappa + s + \frac{d-4}{2})(\kappa + s + \frac{d-6}{2})}  \partial^{b_1}\partial^{b_2}\phi^{b_1b_2a_1\ldots a_{s-2}} \Box^{\kappa-2} \partial^{c_1}\partial^{c_2}\phi^{c_1c_2a_1\ldots a_{s-2}} + \ldots \,.\qquad
\eeq

The following remarks are in order.

\noindent \ibf) For the short and partial-short conformal fields given in the Table, Lagrangian \rf{24022016-05-man} is invariant under gauge transformations
\beq
\label{24022016-10a1-man} && \hspace{-2cm} \delta \phi^{a_1 \ldots a_s}  = \Pi^\tr \partial^{(a_1} \ldots \partial^{a_{t+1}} \xi^{a_{t+2}\ldots a_s)}\,, \qquad t \equiv s + \frac{d-4}{2} - \kappa\,,
\nonumber\\[-5pt]
&& \hspace{4cm}\hbox{ for short and partial-short conformal fields},
\eeq
where a gauge transformation parameter $\xi^{a_{t+2}\ldots a_s}$ is a rank-$(s-1-t)$ traceless totally symmetric tensor field of the Lorentz algebra $so(d-1,1)$ and we use a projector $\Pi^\tr$ to respect the tracelessness constraint \rf{24022016-04-man}.

\noindent \iibf)  For the long and special conformal fields, minimal Lagrangian \rf{24022016-05-man} is gauge variant.

\noindent \iiibf) With the exception of the particular case $\kappa=1$, minimal Lagrangian \rf{24022016-05-man} involves higher-derivatives.

\noindent \ivbf) Lagrangian \rf{24022016-05-man} is invariant under the conformal algebra transformations presented in \rf{02032016-02-man}-\rf{02032016-05-man}. We note then that a realization of spin operator $M^{ab}$ \rf{02032016-07-man} on a space of the traceless tensor field $\phi^{a_1\ldots a_s}$ of the Lorentz algebra $so(d-1,1)$ is well known. Realization of the operator $\Delta$ on a space of the traceless field $\phi^{a_1\ldots a_s}$ is given by \rf{24022016-01-man}. We note also that the operator $R^a$ is  trivially realized on a space of the traceless field $\phi^{a_1\ldots a_s}$, i.e., $R^a=0$.

We now discuss minimal Lagrangian \rf{24022016-05-man} for some particular values of $\kappa$, $s$, and $d$.

\medskip
\noindent {\bf Conformal spin-$s$ field with arbitrary integer $s$ and $\kappa = s+ \frac{d-4}{2}$, $d$-even}. According to our Table such conformal field is referred to as short conformal field. For the short conformal field, Lagrangian \rf{24022016-05-man} takes the form
\beq
\label{24022016-11-man} \LL  & = &  \half  \sum_{n=0}^s
\frac{2^n  (s+\frac{d-2}{2}-n)_n}{  n! (s-n)!(2s +  d-3-n)_n }  (\partial^n\phi)^{a_1\ldots a_{s-n}} \Box^{s+\frac{d-4}{2}-n}(\partial^n\phi)^{a_1\ldots a_{s-n}}\,,
\eeq
where we use the notation as in \rf{24022016-06-man}, \rf{24022016-08-man}.  Lagrangian \rf{24022016-11-man} is invariant under gauge transformations given in \rf{24022016-10a1-man} with $t=0$.
Representation for the minimal Lagrangian of the short conformal field given in \rf{24022016-11-man} was obtained in Ref.\cite{Metsaev:2009ym}. Alternative representations for the minimal Lagrangian of the short conformal field may be found in Refs.\cite{Fradkin:1985am,Segal:2002gd}.

\medskip
\noindent {\bf Conformal spin-$s$ field with arbitrary integer $s$ and $\kappa=1$}. For arbitrary integer $s$ and $\kappa=1$, Lagrangian \rf{24022016-05-man} takes the form
\be \label{24022016-12-man}
\LL  = \frac{1}{2 s!} \phi^{a_1 \ldots a_s} \Box \phi^{a_1 \ldots a_s}
+ \frac{1}{(s-1)!(s + \frac{d-2}{2})}  \partial^b\phi^{ba_1 \ldots a_{s-1}} \partial^c\phi^{ca_1 \ldots a_{s-1}}\,.
\ee
For $d=4$ and $d \geq 4$, Lagrangian \rf{24022016-12-man} with arbitrary $s$ was obtained in the respective Ref.\cite{Iorio:1996ad} and Ref.\cite{Erdmenger:1997wy}.
According to our Table, for $s\geq 2$, $d = 4$, Lagrangian \rf{24022016-12-man} describes the type II partial-short  conformal fields and is invariant under gauge transformations given in \rf{24022016-10a1-man} with $t=s-1$. For  $s\geq 2$, $d=3$ and $s\geq 1$, $d \geq 5$,  Lagrangian \rf{24022016-12-man} describes the special conformal fields and is gauge variant. For  $s=1$, $d=3$, Lagrangian \rf{24022016-12-man} describes the long conformal field and is gauge variant. For $s=1$, $d=4$, Lagrangian \rf{24022016-12-man} describes the short conformal (Maxwell) field.

\noindent {\bf Conformal spin-1 field with arbitrary integer $\kappa\geq 1$}. For this case, Lagrangian \rf{24022016-05-man} takes the following form

\beq
\label{24022016-14-man}  \LL  & = & \frac{1}{2} \phi^a \Box^\kappa \phi^a +  \frac{2\kappa }{2\kappa  + d-2}  \partial^a\phi^a \Box^{\kappa-1} \partial^b\phi^b\,, \qquad  \kappa=1,2,\ldots, \infty\,.
\eeq
Alternatively, Lagrangian \rf{24022016-14-man} can be represented the form similar to the Proca, Lagrangian
\beq
\label{24022016-16-man} g^2 \LL  & = & - \frac{1}{4} F^{ab} \Box^{\kappa-1} F^{ab} - \half m^2 \phi^a \Box^\kappa \phi^a\,,
\\
\label{24022016-17-man} &&  F^{ab} \equiv \partial^a \phi^b - \partial^b \phi^a\,,
\\
\label{24022016-18-man} && g^2 \equiv \frac{2\kappa+d-2}{4\kappa}\,,
\\
\label{24022016-19-man} && m^2 \equiv \frac{2\kappa-d+2}{4\kappa}\,,
\\
\label{24022016-20 -man} && g^2 + m^2 =1\,,
\eeq
where we introduce formally coupling constant $g$ \rf{24022016-18-man} and dimensionless mass parameter $m$ \rf{24022016-19-man}. For $\kappa=1,\ldots, (d-4)/2$, $d\geq 6$, we have $m^2\ne 0$ and Lagrangian \rf{24022016-16-man} is gauge variant. According to our classification in the Table, for $\kappa=1$ and $d\geq 6$, we refer to the field $\phi^a$ as the special conformal field, while, for $\kappa=2,\ldots, (d-4)/2$ and $d\geq 8$, the field $\phi^a$ is referred to as the secondary long conformal field. For $\kappa=(d-2)/2$, we get $m^2=0$ and this case corresponds to a spin-1 short conformal field with gauge invariant Lagrangian  in \rf{24022016-16-man}. For $\kappa > s-2 + [d/2]$ and $d\geq 3$, the Lagrangian given in \rf{24022016-16-man} is gauge variant and the field $\phi^a$ is referred to as the long conformal field.

\medskip
\noindent {\bf Conformal spin-2 field with arbitrary integer $\kappa\geq 1$}. For this case, Lagrangian \rf{24022016-05-man} takes the form

\beq
\label{24022016-21-man} \LL  & = & \frac{1}{4} \phi^{ab} \Box^\kappa \phi^{ab} + \frac{2\kappa }{2\kappa + d}  \partial^b \phi^{ba} \Box^{\kappa-1} \partial^c\phi^{ca}
\nonumber\\
& + & \frac{4\kappa(\kappa-1) }{(2\kappa + d)(2\kappa + d-2)}  \partial^a\partial^b\phi^{ab} \Box^{\kappa-2} \partial^c\partial^e\phi^{ce}\,,\qquad \kappa=1,2,\ldots, \infty\,.\qquad
\eeq
We note that, for $\kappa=2$, $d=4$, Lagrangian \rf{24022016-21-man} describes the Weyl graviton (spin-2 short conformal field) and is invariant under gauge transformations given in \rf{24022016-10a1-man} with $t=0$ and $s=2$.

\noindent {\bf Conformal spin-2 field with $\kappa=\half(d-2)$ and even $d\geq 4$}. For this case, Lagrangian \rf{24022016-05-man} takes the form
\be \label{24022016-22-man}
\LL  =  \frac{1}{4} \phi^{ab} \Box^{\frac{d-2}{2}} \phi^{ab} + \frac{d-2}{2(d-1)}  \partial^b \phi^{ba} \Box^{\frac{d-4}{2}} \partial^c\phi^{ca}  +  \frac{d-4}{4(d-1)}  \partial^a\partial^b\phi^{ab} \Box^{\frac{d-6}{2}} \partial^c\partial^e\phi^{ce}\,.\qquad
\ee
Lagrangian \rf{24022016-22-man} describes the partial-short conformal field corresponding to the value $\kappa=\half(d-2)$ in the Table. For the case of $d=4$, Lagrangian \rf{24022016-22-man} takes the form
\be \label{24022016-23-man}
\LL  =  \frac{1}{4} \phi^{ab} \Box \phi^{ab} + \frac{1}{3}  \partial^b \phi^{ba}  \partial^c\phi^{ca}\,.
\ee
Lagrangian \rf{24022016-23-man} was obtained in Ref.\cite{Barut:1982nj} (see also Refs.\cite{Drew:1980yk}).
Invariance of Lagrangian \rf{24022016-23-man} under gauge transformation \rf{24022016-10a1-man} with $t=1$ was discovered in Ref.\cite{Deser:1983tm}. In Ref.\cite{Bekaert:2013zya}, the spin-2 conformal field described by Lagrangian \rf{24022016-23-man} has been identified with a boundary value of the spin-2 partial-massless field in $AdS_5$
(for arbitrary $s$, $d$, and $t$ see Refs.\cite{Bekaert:2013zya,Barnich:2015tma}).

\medskip
\noindent {\bf Minimal Lagrangian of conformal field from AdS/CFT correspondence}. Minimal Lagrangian \rf{24022016-05-man} can be obtained in the framework of AdS/CFT correspondence. We recall that, in the framework of AdS/CFT correspondence, conformal field that propagates in $R^{d-1,1}$ and has conformal dimension as in \rf{24022016-01-man} is dual to a non-normalizable mode of bulk field that propagates in $AdS_{d+1}$ and has lowest eigenvalue of an energy operator equal to $E_0 = \kappa + \frac{d}{2}$. Let us refer to an action of AdS field evaluated on a solution of the Dirichlet problem as effective action. We will denote the effective action as $S_\eff$. For arbitrary values of $\kappa$, $s$, and $d$, the effective action for spin-$s$ field in $AdS_{d+1}$ was found in Ref.\cite{Metsaev:2011uy} and is given by
\be \label{01042016-01-man}
-S_\eff  =  \frac{\kappa(2\kappa+2s+d-2)}{s!(2\kappa+d-2)} c_\kappa \Gamma^{{\rm stand}} \,, \qquad
\quad c_\kappa \equiv \frac{\Gamma(\kappa+\frac{d}{2})}{\pi^{d/2} \Gamma(\kappa)} \,, \hspace{2cm}
\ee
where a 2-point function $\Gamma^{{\rm stand}}$ appearing in \rf{01042016-01-man} is defined by the relations
\beq
\label{01042016-02-man} && \Gamma^\stand =  \int d^dx_1 d^d x_2\, \Gamma_{12}^\stand\,,
\\
\label{01042016-03-man} &&   \Gamma_{12}^\stand = \phi^{a_1 \ldots a_s
}(x_1) \frac{O_{12}^{a_1b_1} \ldots O_{12}^{a_s b_s}}{|x_{12}|^{2\kappa+d}}
\phi^{b_1\ldots b_s}(x_2) \,,\qquad  O_{12}^{ab}  \equiv \eta^{ab} -
\frac{2x_{12}^a x_{12}^b}{|x_{12}|^2}\,.\qquad
\eeq
From \rf{01042016-02-man}, \rf{01042016-03-man}, we see that $\Gamma^{{\rm stand}}$ is a standard 2-point CFT function for a boundary shadow field $\phi^{a_1 \ldots a_s}$ which has  the conformal dimension given in \rf{24022016-01-man}.
For massless and massive spin-$1$ and spin-2 fields, the normalization factor appearing in front of $\Gamma^{{\rm stand}}$ in \rf{01042016-01-man} is in agreement with the results obtained in the earlier literature (see Refs.\cite{Liu:1998ty,Freedman:1998tz})

For integer values of $\kappa$, the 2-point function \rf{01042016-03-man} is not well defined (see, e.g., Ref.\cite{Aref'eva:2014mia}). Using the regularization $ \kappa \rightarrow \kappa - \epsilon$, $\epsilon \sim 0$,  and the well known expression for UV divergence of the regularized kernel entering
2-point function $\Gamma_{12}^\stand$ \rf{01042016-03-man}%
\footnote{ Useful discussion of technical details of the regularization procedure \rf{01042016-04-man} may be found in Appendix G in Ref.\cite{Alkalaev:2012ic}.}
\beq
\label{01042016-04-man} && \frac{1}{|x|^{2\kappa+d}}\,\,\, \stackrel{\epsilon \sim
0}{\mbox{\Large$\sim$}}\,\,\, \frac{1}{\epsilon} \varrho_\kappa \Box^\kappa
\delta^d(x)\,,  \qquad  \varrho_\kappa \equiv \frac{\pi^{d/2}}{4^\kappa \Gamma(\kappa
+ 1)\Gamma(\kappa + \frac{d}{2})}\,,
\eeq
we verify that UV divergence of the 2-point function $\Gamma^\stand$ \rf{01042016-02-man} takes the form
\be \label{01042016-05-man}
\Gamma^\stand\Bigl|_{\epsilon \sim 0}  \,\,\, {\mbox{\Large$\sim$}}\,\,\,   \frac{2 s!}{\epsilon} \frac{2\kappa + d -2}{2\kappa + 2s+ d -2} \varrho_\kappa  S_\min\,, \qquad S_\min = \int d^dx    \LL\,,
\ee
where Lagrangian appearing in \rf{01042016-05-man} is nothing but the minimal Lagrangian given in \rf{24022016-05-man}. Plugging \rf{01042016-05-man} into \rf{01042016-01-man} we see that the UV divergence of the effective action is proportional to the minimal action of conformal field
\be \label{01042016-06-man}
- S_\eff\Bigl|_{\epsilon \sim 0}  \,\,\, {\mbox{\Large$\sim$}}\,\,\, \frac{1}{\epsilon} \frac{2}{ 4^\kappa\Gamma^2(\kappa)} S_\min \,.
\ee
From relation \rf{01042016-06-man}, we see that UV divergence of the effective action for field in $AdS_{d+1}$ with integer value of $\kappa$ is indeed realized as the minimal action for conformal field in $R^{d-1,1}$.

%%%%%%%%%%%%%%%%%%%%%%%%%%%%%%%%%%%%%%%%%%%%%%%%%%%%%%%%%%%%%%%%%%%%%%%%%%%%%%%%%%%%%%%%%%%%%
%%%%%%%%%%%%%%%%%%%%%%%%%%%%%%%%%%%%%%%%%%%%%%%%%%%%%%%%%%%%%%%%%%%%%%%%%%%%%%%%%%%%%%%%%%%%%
\newsection{ \large Ordinary-derivative gauge
invariant Lagrangian of conformal field}\label{sec-04}
%%%%%%%%%%%%%%%%%%%%%%%%%%%%%%%%%%%%%%%%%%%%%%%%%%%%%%%%%%%%%%%%%%%%%%%%%%%%%%%%%%%%%%%%%%%%%
%%%%%%%%%%%%%%%%%%%%%%%%%%%%%%%%%%%%%%%%%%%%%%%%%%%%%%%%%%%%%%%%%%%%%%%%%%%%%%%%%%%%%%%%%%%%%

{\bf Field content for long, partial-short and special conformal fields}. In order to develop an ordinary-derivative gauge invariant metric-like formulation of totally symmetric arbitrary spin-$s$ conformal field that propagates in $R^{d-1,1}$ and has conformal dimension given in \rf{24022016-01-man}, we introduce the following set of real-valued scalar,
vector, and tensor fields of the Lorentz algebra $so(d-1,1)$:
\be
\label{26022016-00-man}  \phi_{\lambda,k'}^{a_1\ldots a_{s'}}(x)\,,
\ee
where labels $s'$, $\lambda$, $k'$ take the following values

{\small
\beq
\label{26022016-01-man} && \hspace{-1.5cm} s'=0,1,\ldots,s\,, \hspace{0.6cm}  \lambda\in [s-s']_2 \,,  \hspace{0.6cm}     k'\in [\kappa-1+\lambda]_2\,, \hspace{0.7cm} \kappa -1 +\lambda \geq 0,
\nonumber\\[-2pt]
&& \hspace{6.5cm} \hbox{ for long and secondary long conformal fields};
\\[15pt]
\label{26022016-02-man} && \hspace{-1.5cm} s'=0,1,\ldots,s\,, \hspace{0.7cm}    \lambda\in [s-s']_2 \,, \hspace{0.6cm}    k'\in [\kappa-1+\lambda]_2\,, \hspace{0.7cm} \kappa -1 +\lambda \geq 0,
\nonumber\\[-2pt]
&&  \hspace{2cm} s-s'+2-2\kappa  \leq \lambda   \,,
\nonumber\\[-2pt]
&& \hspace{6.5cm} \hbox{ for special conformal fields};
\\[15pt]
\label{26022016-03-man} && \hspace{-1.5cm} s'=0,1,\ldots, s\,, \qquad  \lambda \in [s-s']_2 \,,  \qquad  k'\in [\kappa-1+\lambda]_2\,, \hspace{0.5cm} \kappa -1 +\lambda \geq 0,
\nonumber\\[-2pt]
&&  \hspace{2cm}   \lambda \leq s + s' +d-4-2\kappa \,,
\nonumber\\
&& \hspace{6.5cm} \hbox{ for type I partial-short conformal fields};
\\[15pt]
\label{26022016-04-man} && \hspace{-1.5cm} s'=0,1,\ldots, s\,, \qquad  \lambda \in [s-s']_2 \,,  \qquad  k'\in [\kappa-1+\lambda]_2\,, \hspace{0.5cm} \kappa -1 +\lambda \geq 0,
\nonumber\\[-2pt]
&&  \hspace{2cm} s-s'+2-2\kappa  \leq \lambda \leq s + s' +d-4-2\kappa \,,
\nonumber\\
&& \hspace{6.5cm} \hbox{ for type II partial-short conformal fields};
\\[15pt]
\label{26022016-05-man} && \hspace{-1.5cm} s'=0,1,\ldots, s\,, \qquad  \lambda  =s'-s \,,  \hspace{1.2cm}  k'\in [k_{s'}]_2\,, \hspace{1.5cm} k_{s'}\geq 0\,, \hspace{0.7cm} k_{s'}\equiv s'+\frac{d-6}{2},
\nonumber\\
&& \hspace{6.5cm} \hbox{ for short conformal fields}.
\eeq
}
Relation $p \in [q]_2$ appearing in \rf{26022016-01-man}-\rf{26022016-05-man} is defined in \rf{29022016-05a1-man}. We recall also that values of $s$ and $\kappa$ for the various conformal fields are defined in the Table.

The following remarks are in order.

\noindent \ibf) In the catalogue \rf{26022016-00-man}-\rf{26022016-05-man}, fields $\phi_{\lambda,k'}^{a_1\ldots a_{s'}}$ with $s'=0$ and $s'=1$ are the respective scalar and vector fields of the Lorentz algebra $so(d-1,1)$,
while fields $\phi_{\lambda,k'}^{a_1\ldots a_{s'}}$ with $s'>1$ are totally symmetric rank-$s'$ tensor fields of the Lorentz algebra. By definition, the tensor fields $\phi_{\lambda,k'}^{a_1\ldots a_{s'}}$ with $s'\geq 4$ are double traceless tensor fields,
\be  \label{26022016-04a1-man}
\phi_{\lambda,k'}^{aabba_5\ldots a_{s'}}=0\,, \qquad
s'\geq 4\,.
\ee

\noindent \iibf) Conformal dimensions of the fields $\phi_{\lambda,k'}^{a_1\ldots
a_{s'}}$ \rf{26022016-00-man} are given by the relation
\be \label{26022016-05a1-man}
\Delta(\phi_{\lambda,k'}^{a_1\ldots a_{s'}}) = \frac{d-2}{2} + k'\,.
\ee

\noindent \iiibf) Taking into account the restrictions on the label $\lambda$ given in the second lines in  \rf{26022016-02-man}-\rf{26022016-04-man}, we see that the domains of values of the label $\lambda$ for the special and partial-short fields \rf{26022016-02-man}-\rf{26022016-04-man} are obtained by decreasing the domain of values of the label $\lambda$ for the long fields in \rf{26022016-01-man}. In other words, the field contents of the special, type I partial-short and type II partial-short conformal fields are obtained from the field content of the long conformal fields by setting to zero those fields in \rf{26022016-01-man} whose values of $\lambda$ do not respect constraints in the second lines in the respective relations \rf{26022016-02-man}, \rf{26022016-03-man}, and \rf{26022016-04-man}. It is the restrictions on $\lambda$ appearing in the second lines in  \rf{26022016-02-man}-\rf{26022016-04-man} and AdS/CFT dictionary that motivate us to classify fields into the special, types I and II partial-short conformal fields (see below). Also we see that the field content of the short conformal field \rf{26022016-05-man} is obtained by setting to zero all fields in \rf{26022016-01-man} with $\lambda \ne s'-s$. The field content entering the ordinary-derivative formulation of the short conformal fields has been found in Ref.\cite{Metsaev:2007rw}.

\vspace{5pt}
\noindent \ivbf) The terminology we use in this paper is inspired by AdS/CFT dictionary. Namely, a conformal field that propagates in $R^{d-1,1}$ and has conformal dimension as in \rf{24022016-01-man} is dual to a non-normalizable mode of bulk field that propagates in $AdS_{d+1}$ and has lowest eigenvalue of an energy operator equal to $E_0 = \kappa + \frac{d}{2}$.
Our long and secondary long conformal fields \rf{26022016-01-man} are dual to AdS massive fields associated with the respective unitary and non-unitary irreps of the $so(d,2)$ algebra. Conformal field in \rf{26022016-02-man} is related to AdS massive field associated with non-unitary irrep of the $so(d,2)$ algebra. In view of the restriction on $\lambda$ in the second line in \rf{26022016-02-man} we refer to such conformal field as special conformal field.
Conformal fields in \rf{26022016-03-man}, \rf{26022016-04-man} are dual to AdS partial-massless fields.%
\footnote{ Arbitrary spin partial-massless fields in $AdS_4$ were first studied in Refs.\cite{Deser:2001xr}. Generalization of results in the latter references to $AdS_{d+1}$, $d\geq 3$, may be found in Refs.\cite{Zinoviev:2001dt,Metsaev:2006zy}. Discussion of various aspects of free and interacting partial-massless AdS fields may be found in the respective Ref.\cite{Skvortsov:2006at} and Ref.\cite{Joung:2014aba}.}
In view of the restrictions on $\lambda$ in the second lines in \rf{26022016-03-man} and \rf{26022016-04-man} we  refer to such conformal fields as the respective type I and type II partial-short conformal fields. Conformal field in \rf{26022016-05-man} is dual to massless AdS field. Therefore we refer to such conformal field as short conformal field.

\medskip
\noindent \vbf) For the special and partial-short conformal fields, the restriction $\kappa-1+\lambda\geq 0$ appearing in \rf{26022016-01-man}-\rf{26022016-04-man} is satisfied automatically, while, for some long conformal field in $R^{2,1}$, this restriction leads to a constraint on the field content. Namely, using expression for $\kappa$ corresponding to the long conformal field in $R^{2,1}$ with $N=0$ (see the Table), we find the relation $\kappa=s$ and note that, for $s'=0$ and $\lambda=-s$, the restriction $\kappa-1+\lambda\geq 0$ is not satisfied. This implies that the fields with $s'=0$ (scalar fields) and $\lambda=-s$ do not enter the field content of the spin-$s$ long conformal field in $R^{2,1}$ that has $\kappa=s$, (the case of $N=0$ in the Table). We note then that, for a short conformal field in $R^{3,1}$, the restriction $k_{s'}\geq 0$ appearing in \rf{26022016-05-man} also leads to some constraint on the field content. Namely, using expression for $k_{s'}$ in \rf{26022016-05-man}, we see that for $d=4$ and $s'=0$, the constraint $k_{s'}\geq 0$ is not satisfied. This implies that fields with $s'=0$ (scalar fields) do not enter the field content of the ordinary-derivative formulation of the short conformal field in $R^{3,1}$.

\medskip
\noindent \vibf) For $d=4$, it is easy to see that the restrictions on $\lambda$ in the second line in \rf{26022016-04-man} lead to the restriction $s'\geq 1$. This implies scalar fields do not enter field content of the ordinary-derivative formulation of the type II partial-short conformal fields in $R^{3,1}$.

\medskip

\medskip
\noindent {\bf Generating form of field content}. In order to obtain a gauge invariant description of a conformal field in an easy--to--use form, we use oscillators $\alpha^a$, $\alpha^z$, $\zeta$, $\alpha^\oplussm$, $\alpha^\ominussm$ and
collect all fields given in \rf{26022016-01-man}-\rf{26022016-05-man} into the following ket-vector:
\be  \label{26022016-06-man}
\phik =  \sum_{s', \lambda, k'}  |\phi_{\lambda,k'}^{s'}\rangle\,,
\ee
where basis ket-vectors $|\phi_{\lambda,k'}^{s'}\rangle$ appearing in \rf{26022016-06-man} take the form
\beq
\label{26022016-07-man} |\phi_{\lambda,k'}^{s'}\rangle & \equiv  &
\frac{\zeta^{\half(s-s'+\lambda)}\alpha_z^{\half(s-s'-\lambda)}}{\sqrt{(\frac{s-s'+\lambda}{2})!(\frac{s-s'-\lambda}{2})! } }
\frac{(\alpha^\oplussm)^{\half(\kappa-1+\lambda-k')} (\alpha^\ominussm)^{\half(\kappa-1+\lambda+k')}}{(\frac{\kappa-1+\lambda+k'}{2})! \, s'! }
\nonumber\\
& \times & \alpha^{a_1}\ldots \alpha^{a_{s'}}\phi_{\lambda,k'}^{a_1\ldots a_{s'}}(x)|0\rangle\,,
\eeq
and, depending on the type of the conformal field, the summation indices $s'$, $\lambda$, $k'$ in \rf{26022016-06-man} run over values given in \rf{26022016-01-man}-\rf{26022016-05-man}. Note that, for the short conformal field, the $\lambda$ is fixed (see \rf{26022016-05-man}).

From relations \rf{26022016-06-man},\rf{26022016-07-man},  it is easy to see that the ket-vector
$\phik$ satisfies the following algebraic constraints
\beq
\label{26022016-08-man} && (N_\alpha + N_z + N_\zeta - s)\phik =0 \,,
\\
\label{26022016-09-man} && (N_z - N_\zeta + N_{\alpha^\oplussm}+ N_{\alpha^\ominussm} - \kappa+1)\phik =0 \,,
\eeq
while the basis ket-vectors $|\phi_{\lambda,k'}^{s'}\rangle$ \rf{26022016-07-man} satisfy the algebraic constraints
\beq
\label{26022016-10-man} && N_\alpha |\phi_{\lambda,k'}^{s'}\rangle  \hspace{1.7cm} = \,  s'|\phi_{\lambda,k'}^{s'}\rangle\,,
\\
\label{26022016-11-man} && (N_\zeta - N_z)|\phi_{\lambda,k'}^{s'}\rangle \hspace{0.3cm} =\, \lambda |\phi_{\lambda,k'}^{s'}\rangle\,,
\\
\label{26022016-12-man} && (N_{\alpha^\ominussm}  - N_{\alpha^\oplussm})|\phi_{\lambda,k'}^{s'}\rangle =\,  k'|\phi_{\lambda,k'}^{s'}\rangle \,,
\eeq
where a definition of the operators $N_\alpha$, $N_\zeta$, etc., may be found in relations \rf{29022016-08-man}, \rf{29022016-09-man}.

From relation \rf{26022016-08-man}, we learn that the ket-vector $\phik$ is a
degree-$s$ homogeneous polynomial in the oscillators $\alpha^a$,
$\alpha^z$, $\zeta$.  Relation \rf{26022016-10-man} tells us that the basis ket-vector $|\phi_{\lambda,k'}^{s'}\rangle$ is a degree-$s'$  homogeneous polynomial in the oscillators $\alpha^a$. From relations \rf{26022016-11-man} and \rf{26022016-12-man} we learn that the ket-vector $|\phi_{\lambda,k'}^{s'}\rangle$ is an eigenvector of the respective operators $N_\zeta - N_z$ and $N_{\alpha^\ominussm}  - N_{\alpha^\oplussm}$.
Also we note that, in terms of the ket-vector $\phik$, constraint given in  \rf{26022016-04a1-man} takes the following form
\be \label{26022016-14-man}
(\bar{\alpha}^2)^2 \phik = 0 \,.
\ee

\noindent {\bf Remark on the short conformal fields}. As we have already said, the field content entering the ordinary-derivative formulation of the short conformal fields given in \rf{26022016-00-man}, \rf{26022016-05-man} has been found in Ref.\cite{Metsaev:2007rw}. For the reader's convenience and in order to match result in Ref.\cite{Metsaev:2007rw} with the presentation in this paper we now write down the explicit form of the ket-vector $\phik$ for the short conformal field. Such explicit form is obtained by plugging $\kappa=s+\frac{d-4}{2}$ and the labels $s'$, $\lambda$, $k'$ given in \rf{26022016-05-man} into ket-vector \rf{26022016-06-man}, \rf{26022016-07-man}. Doing so, we get
\beq
\label{27022016-03-man} && \hspace{-2.5cm}  \phik  =  \sum_{s' k'}
|\phi_{s'-s,k'}^{s'}\rangle \,,  \hspace{1.8cm}  \hbox{ for short conformal fields},
\\[5pt]
\label{27022016-04-man} |\phi_{s'-s,k'}^{s'}\rangle & \equiv  &
\frac{ \alpha_z^{ s-s' } (\alpha^\oplussm)^{\half(k_{s'}-k')} (\alpha^\ominussm)^{\half(k_{s'}+k')} }{ \sqrt{( s-s' )!}\,\, (\frac{k_{s'}+k'}{2})!\, s'!}
\alpha^{a_1}\ldots \alpha^{a_{s'}}\phi_{s'-s,k'}^{a_1\ldots a_{s'}}(x)|0\rangle\,,
\eeq
where $k_{s'}$ is defined in \rf{26022016-05-man}.

Comparing the basis ket-vectors in \rf{26022016-07-man} and \rf{27022016-04-man}, we see that the basis ket-vectors of the short conformal field \rf{27022016-04-man} do not depend on the oscillator $\zeta$.%
\footnote{ We note the clash of the notation and conventions in this paper and in Ref.\cite{Metsaev:2007rw}. Namely, the fields $\phi_{s'-s,k'}^{a_1\ldots a_{s'}}$ appearing in \rf{27022016-03-man}, \rf{27022016-04-man} are denoted by $\phi_{k'}^{a_1\ldots a_{s'}}$  in Ref.\cite{Metsaev:2007rw}. Also we note that the ket-vector $\phik$ of the short conformal field in Ref.\cite{Metsaev:2007rw} is obtained from \rf{27022016-03-man}, \rf{27022016-04-man} by using the replacements $\alpha^z \rightarrow \zeta$, $\alpha^\oplussm \rightarrow \upsilon^\ominussm$, $\alpha^\ominussm \rightarrow \upsilon^\oplussm$. }
This implies that ket-vector $\phik$ \rf{27022016-03-man} satisfies the following algebraic constraints
\beq
\label{27022016-04b1-man} && (N_\alpha + N_z  - s)\phik =0 \,,
\\[-5pt]
\label{27022016-04b2-man} && (N_z  + N_{\alpha^\oplussm}+ N_{\alpha^\ominussm} - k_s )\phik =0 \,,\qquad k_s \equiv s + \frac{d-6}{2}\,,
\eeq
while basis ket-vectors $|\phi_{s'-s,k'}^{s'}\rangle$ \rf{27022016-04-man} satisfy the algebraic constraints
\beq
\label{27022016-04b3-man} && N_\alpha |\phi_{s'-s,k'}^{s'}\rangle  \hspace{1.9cm} = \,  s'|\phi_{s'-s,k'}^{s'}\rangle\,,
\\
\label{27022016-04b4-man} &&  N_z|\phi_{s'-s,k'}^{s'}\rangle \hspace{1.7cm} =\, (s-s') |\phi_{s'-s,k'}^{s'}\rangle\,,
\\
\label{27022016-04b5-man} && (N_{\alpha^\ominussm}  - N_{\alpha^\oplussm})|\phi_{s'-s,k'}^{s'}\rangle \, =\,  k'|\phi_{s'-s,k'}^{s'}\rangle \,.
\eeq

\noindent{\bf Gauge invariant Lagrangian}. We now discuss an ordinary-derivative gauge invariant Lagrangian for all conformal fields given in the Table. We find the following representation for
action and ordinary-derivative Lagrangian in terms of the ket-vector $\phik$ above discussed:
\beq
\label{28022016-01-man} && \hspace{-1.3cm} S = \int d^d x\, \LL\,,
\\
\label{28022016-02-man} \LL  & = & \frac{1}{2}
\phibr (1-\frac{1}{4}\alpha^2\alphab^2)
(\Box - M^2)
\phik + \half
\langle \Lb
\phi| \Lb\phi\rangle \,,\qquad
\\
\label{28022016-03-man} && M^2 \equiv \alpha^\oplussm  \bar\alpha^\oplussm\,,\qquad
\\
\label{28022016-04-man} &&   \Lb \equiv \albpar
- \half \alpar \bar\alpha^2 - \Pi^\smponetwo \eb_1
+ \half e_1 \bar\alpha^2\,,\qquad
\\
\label{28022016-05-man} && \qquad  e_1 = \zeta e_\zeta  \alpha^\oplussm + \alpha^z e_z \alphab^\oplussm\,,
\\
\label{28022016-06-man} && \qquad \eb_1 = - \alphab^\oplussm \eb_\zeta \zetab  - \alpha^\oplussm e_z \alphab^z \,,
\eeq
where operators $\Box$, $\alpha^2$, $\alpar$, etc., appearing in \rf{28022016-02-man}, \rf{28022016-04-man} are defined in \rf{29022016-06-man}-\rf{29022016-10-man}.
We now describe quantities entering the Lagrangian in \rf{28022016-02-man}.

\medskip
\noindent \ibf) Bra-vector $\phibr$ entering Lagrangian \rf{28022016-02-man} is defined according the rule
\be \label{28022016-07-man}
\phibr \equiv (\phik)^\dagger \betabf\,,
\ee
where an operator $\betabf$ appearing in \rf{28022016-07-man} takes the following form for various conformal fields:
\beq
\label{28022016-08-man} && \betabf  = 1\,,  \hspace{1.8cm} \hbox{for long conformal fields in } R^{d-1,1}  \hbox{ and}
\nonumber\\[-7pt]
&& \hspace{3.1cm}  \hbox{for short conformal fields in } R^{d-1,1},  \hbox{ $d$-even}; \qquad
\\
\label{28022016-09-man} && \betabf =  e^{\irm \pi N_\zeta},   \hspace{1.2cm}  \hbox{for secondary-long, special, and partial-short  fields in } R^{d-1,1},  \hbox{ $d$-even}; \qquad
\nonumber\\[-3pt]
&&
\\[-7pt]
\label{28022016-10-man}&& \betabf = e^{\irm \pi N_\zeta \epsilon(N_\zeta - t -1) + \irm\pi (t +\half) \theta(N_\zeta - t -1)}\,, \qquad t \equiv s+ \frac{d-4}{2} - \kappa\,,
\nonumber\\[-3pt]
&& \hspace{3.1cm} \hbox{for secondary long and  special fields in } R^{d-1,1}, \hbox{ $d$-odd};
\eeq
where symbols $\epsilon$ and $\theta$ appearing in \rf{28022016-10-man} are defined as
\beq
&& \epsilon(n) = 1 \hbox{ for } n < 0, \qquad \epsilon(n) = 0  \hbox{ for } n\geq 0\,,
\\
&& \theta(n) = 0 \hbox{ for } n < 0, \qquad \theta(n) = 1  \hbox{ for } n \geq 0\,.
\eeq

Note that, on a space of the ket-vector $\phik$, one has the relation $\betabf^2=1$. Appearance of the operator $\betabf$ in \rf{28022016-07-man} is related to the fact that our Lagrangian is constructed out of real-valued fields of the Lorentz algebra $so(d-1,1)$.
Only for the long and short conformal fields, eigenvalues of the operator $\betabf$ are strictly positive \rf{28022016-08-man}. For the secondary long, special, and partial-short conformal fields,
we see that, depending on eigenvalues of the operator $N_\zeta$ on a space of ket-vector $\phik$ \rf{26022016-06-man}, a spectrum of the operator $\betabf$ involves both the positive and negative eigenvalues. For the case of the long and short fields, the strictly positive spectrum of the operator $\betabf$  can intuitively  be explained by the fact that, in the framework of AdS/CFT, the long and short conformal fields in $R^{d-1,1}$ are related to unitary massive and massless fields in $AdS_{d+1}$.
Accordingly, the appearance of the both positive and negative eigenvalues of $\betabf$ for the secondary long, special, and partial-short conformal fields can intuitively be explained by the fact that, in the framework of AdS/CFT, the secondary long and special conformal fields in $R^{d-1,1}$ are related to non-unitary massive fields in $AdS_{d+1}$, while the partial-short conformal fields in $R^{d-1,1}$ are related to partial-massless fields in $AdS_{d+1}$ which are also non-unitary.

\noindent \iibf) Operators $e_\zeta$, $e_z$, $\eb_\zeta$ appearing in \rf{28022016-05-man}, \rf{28022016-06-man} take the following form for the various conformal fields
\beq
\label{28022016-23-man} && \hspace{-1cm} e_\zeta = r_\zeta\,, \hspace{1.1cm} e_z = r_z\,, \qquad \eb_\zeta = r_\zeta\,, \hspace{1.7cm}  \hbox{for long fields in } R^{d-1,1};
\\[12pt]
\label{28022016-24-man} && \hspace{-1cm}  e_\zeta = |r_\zeta| \,, \qquad e_z = r_z\,, \qquad \eb_\zeta = - |r_\zeta|,  \hspace{1.2cm}  \hbox{for secondary long, special, and partial-short}
\nonumber\\
&& \hspace{7.1cm}  \hbox{ fields in } R^{d-1,1}, \hbox{ $d$-even}; \qquad
\\[12pt]
\label{28022016-25-man} && \hspace{-1cm} e_\zeta = |r_\zeta| \,, \qquad e_z = r_z\,, \qquad \eb_\zeta = e^{ \irm \pi \epsilon(N_\zeta - t )} |r_\zeta|\,,
\nonumber\\
&& \hspace{4cm}
\hbox{for secondary long and  special fields in } R^{d-1,1},  \hbox{ $d$-odd};
\\[12pt]
\label{28022016-26-man} && \hspace{-1cm} e_\zeta = 0 \,, \hspace{1.2cm} e_z = \Bigl[\frac{2s+d-4-N_z}{2s+d-4-2N_z}\Bigr]^\half\,, \hspace{1cm}
  \hbox{for short fields in } R^{d-1,1}, \hbox{ $d$-even}; \qquad
\eeq
where operators $r_\zeta$, $r_z$ appearing in \rf{28022016-23-man}-\rf{28022016-25-man} are defined in \rf{28022016-21-man}, \rf{28022016-22-man}, while the parameter $t$ is defined in \rf{28022016-10-man}.

\noindent \iiibf) The quantity $|\Lb\phi\rangle$ appearing in \rf{28022016-02-man} is defined as
$|\Lb\phi\rangle \equiv \Lb \phik$, while
$\langle \Lb\phi|$ is defined as $\langle \Lb\phi| \equiv (|\Lb\phi\rangle)^\dagger \betabf$.
We note that if, in expression for $\Lb$ \rf{28022016-04-man}, we set $e_1=0$, $\eb_1=0$, then the quantity $|\Lb\phi\rangle$ becomes the standard de Donder divergence entering Lagrangian of a massless field in $R^{d-1,1}$.%
\footnote{ Interesting applications of the standard de Donder divergence for studying various aspects of higher-spin field theory may be found in Refs.\cite{Ponomarev:2016jqk}. We think that our modified de Donder gauge could be useful for the computations in conformal higher-spin field theory discussed in Ref.\cite{Joung:2015eny}.}
For this reason, the quantity $|\Lb\phi\rangle$ with $\Lb$ as in \rf{28022016-04-man} we refer to as modified de Donder divergence. Obviously, it is the use of the modified de Donder divergence that allows us to simplify significantly our representation for the gauge invariant Lagrangian given in \rf{28022016-02-man}.

\medskip
\noindent \ivbf)  If, in Lagrangian \rf{28022016-02-man}, we set $M^2=0$, $e_1=0$, $\eb_1=0$, then we are left with two derivative contributions to the Lagrangian. It is easy to see that, for spin-1, spin-2, and spin-$s$, $s>2$,  fields, those two-derivative contributions take the form of the respective Maxwell, Einstein-Hilbert, and Fronsdal kinetic terms.

\noindent \vbf) Using the above-given explicit expressions for the operators $e_1$, $\eb_1$, and $\betabf$, we check that, on a space of the ket-vector $\phik$, the following hermitian conjugation rules
\be
\label{02032016-15-man}   \betabf^\dagger =  \betabf\,, \qquad (\betabf e_1)^\dagger = - \betabf \eb_1
\ee
hold true. For the derivation of relations \rf{02032016-15-man}, we use the fact that the ket-vector $\phik$ satisfies the relation $e^{2\pi \irm N_\zeta}\phik=\phik$ which, in turn, implies the relation $\betabf^2\phik=\phik$.

%%%%%%%%%%%%%%%%%%%%%%%%%%%%%%%%%%%%%%%%%%%%%%%%%%%%%%%%%%%%%%%%%%%%%%%%%
%%%%%%%%%%%%%%%%%%%%%%%%%%%%%%%%%%%%%%%%%%%%%%%%%%%%%%%%%%%%%%%%%%%%%%%%%
\newsection{\large Gauge symmetries of conformal fields in ordinary-derivative approach }\label{sec-05}
%%%%%%%%%%%%%%%%%%%%%%%%%%%%%%%%%%%%%%%%%%%%%%%%%%%%%%%%%%%%%%%%%%%%%%%%%
%%%%%%%%%%%%%%%%%%%%%%%%%%%%%%%%%%%%%%%%%%%%%%%%%%%%%%%%%%%%%%%%%%%%%%%%%

{\bf Gauge transformation parameters for long, partial-short, short, and special conformal fields}. In order to discuss gauge symmetries of the ordinary-derivative Lagrangian \rf{28022016-02-man}, we introduce the following set of scalar,
vector, and tensor gauge transformation parameters:
\be \label{28022016-27a-man}
\xi_{\lambda,k'}^{a_1\ldots a_{s'}}(x)\,,
\ee
where labels $s'$, $\lambda$, $k'$ take the following values

{\small
\beq
\label{28022016-27-man} && \hspace{-1.5cm} s'=0,1,\ldots,s-1\,, \hspace{0.6cm}  \lambda\in [s-1-s']_2 \,,  \hspace{0.6cm}     k'\in [\kappa-1+\lambda]_2\,, \hspace{0.7cm} \kappa -1 +\lambda \geq 0,
\nonumber\\[-2pt]
&& \hspace{6.2cm} \hbox{ for long and secondary long conformal fields};
\\[15pt]
\label{28022016-28-man} && \hspace{-1.5cm} s'=0,1,\ldots,s-1\,, \hspace{0.6cm}    \lambda\in [s-1-s']_2 \,, \hspace{0.6cm}    k'\in [\kappa-1+\lambda]_2\,, \hspace{0.7cm} \kappa -1 +\lambda \geq 0,
\nonumber\\[-2pt]
&&  \hspace{2.5cm} s-s'+1-2\kappa  \leq \lambda   \,,
\nonumber\\[-2pt]
&& \hspace{6.2cm} \hbox{ for special conformal fields};
\\[15pt]
\label{28022016-29-man} && \hspace{-1.5cm} s'=0,1,\ldots, s-1\,, \qquad  \lambda \in [s-1-s']_2 \,,  \qquad  k'\in [\kappa-1+\lambda]_2\,, \hspace{0.5cm} \kappa -1 +\lambda \geq 0,
\nonumber\\[-2pt]
&&  \hspace{2.5cm}   \lambda \leq s + s' +d-3-2\kappa \,,
\nonumber\\
&& \hspace{6.2cm} \hbox{ for type I partial-short conformal fields};
\\[15pt]
\label{28022016-30-man} && \hspace{-1.5cm} s'=0,1,\ldots, s-1\,, \qquad  \lambda \in [s-1-s']_2 \,,  \qquad  k'\in [\kappa-1+\lambda]_2\,, \hspace{0.5cm} \kappa -1 +\lambda \geq 0,
\nonumber\\[-2pt]
&&  \hspace{2.5cm} s-s'+1-2\kappa  \leq \lambda \leq s + s' +d-3-2\kappa \,,
\nonumber\\
&& \hspace{6.2cm} \hbox{ for type II partial-short conformal fields};
\\[15pt]
\label{28022016-31-man} && \hspace{-1.5cm} s'=0,1,\ldots, s-1\,, \qquad  \lambda  =s'+1-s \,,  \hspace{1.2cm}  k'\in [k_{s'}+1]_2\,, \hspace{1.5cm} k_{s'}\geq 0\,, \hspace{0.7cm}
\nonumber\\
&& \hspace{2.5cm} k_{s'}\equiv s'+\frac{d-6}{2}, \hspace{1cm} \hbox{ for short conformal fields};
\eeq
}
and the relation $p \in [q]_2$ appearing in \rf{28022016-27-man}-\rf{28022016-31-man} is defined in \rf{29022016-05a1-man}.  Note also that the values of $s$ and $\kappa$ for the various conformal fields are defined in the Table.

The following remarks are in order.

\noindent \ibf) In the catalogue of gauge transformation parameters \rf{28022016-27-man}-\rf{28022016-31-man}, parameters $\xi_{\lambda,k'}^{a_1\ldots a_{s'}}$ with $s'=0$ and $s'=1$ are the respective scalar and vector fields of the Lorentz algebra $so(d-1,1)$,
while parameters $\xi_{\lambda,k'}^{a_1\ldots a_{s'}}$ with $s'>1$ are totally symmetric rank-$s'$ tensor fields of the Lorentz algebra. By definition, the gauge transformation parameters $\xi_{\lambda,k'}^{a_1\ldots a_{s'}}$ with $s'\geq 2$ are traceless tensor fields,
\be \label{28022016-30a1-man}
\xi_{\lambda,k'}^{aaa_3\ldots a_{s'}}=0\,, \qquad
s'\geq 2\,.
\ee

\noindent \iibf) Conformal dimensions of the parameters $\xi_{\lambda,k'}^{a_1\ldots
a_{s'}}$ are given by the relation
\be \label{28022016-31a1-man}
\Delta(\xi_{\lambda,k'}^{a_1\ldots a_{s'}}) = \frac{d-4}{2} + k'\,.
\ee

\noindent \iiibf) For the reader's convenience, we note two alternative and equivalent simple rules for getting
the values of the labels in \rf{28022016-27-man}-\rf{28022016-31-man}. Namely, values of the labels in \rf{28022016-27-man}-\rf{28022016-31-man} can be obtained by using one of the two replacements for labels $s'$, $s$, and $d$ in \rf{26022016-01-man}-\rf{26022016-05-man} which we present as rule I and rule II,
\beq
&& \hspace{-2.5cm} \hbox{ rule I:} \qquad \  s' \rightarrow s'+1,\hspace{5cm} \hbox{ $s$ and $d$ fixed};
\\
&& \hspace{-2.5cm} \hbox{ rule II:} \qquad  s\rightarrow s-1, \qquad d\rightarrow d + 2\,, \hspace{2.2cm} \hbox{ $s'$ fixed}. \qquad
\eeq

In order to simplify a presentation of gauge symmetries we use the oscillators $\alpha^a$, $\alpha^z$, $\zeta$, $\alpha^\oplussm$, $\alpha^\ominussm$ and collect all gauge transformation parameters given in \rf{28022016-27-man}-\rf{28022016-31-man} into the following ket-vector
\be \label{28022016-32-man}
\xik  =  \sum_{s',\lambda, k' } |\xi_{\lambda,k'}^{s'}\rangle \,,
\ee
where basis ket-vectors appearing in \rf{28022016-32-man} take the form
\beq
\label{28022016-33-man} |\xi_{\lambda,k'}^{s'}\rangle & \equiv  &
\frac{\zeta^{\half(s-1-s'+\lambda)}\alpha_z^{\half(s-1-s'-\lambda)}}{\sqrt{(\frac{s-1-s'+\lambda}{2})!(\frac{s-1-s'-\lambda}{2})! } }
\frac{(\alpha^\oplussm)^{\half(\kappa-1+\lambda-k')} (\alpha^\ominussm)^{\half(\kappa-1+\lambda+k')}}{(\frac{\kappa-1+\lambda+k'}{2})! \, s'! }
\nonumber\\
& \times &  \alpha^{a_1}\ldots \alpha^{a_{s'}}\xi_{\lambda,k'}^{a_1\ldots a_{s'}}(x)|0\rangle
\eeq
and, depending on the type of the conformal field, the summation indices $s', \lambda, k'$ in \rf{28022016-32-man} run over values given in \rf{28022016-27-man}-\rf{28022016-31-man}.

Using relations \rf{28022016-32-man},\rf{28022016-33-man},  it is easy to see that ket-vector
$\xik$ \rf{28022016-32-man} satisfies the following relations
\beq
\label{28022016-34-man} && (N_\alpha + N_z + N_\zeta - s + 1)\xik =0 \,,
\\
\label{28022016-35-man} && (N_z - N_\zeta + N_{\alpha^\oplussm}+ N_{\alpha^\ominussm} - \kappa+1)\xik =0 \,,
\eeq
while the basis ket-vectors $|\xi_{\lambda,k'}^{s'}\rangle$ \rf{28022016-33-man} satisfy the relations
\beq
\label{28022016-36-man} && N_\alpha |\xi_{\lambda,k'}^{s'}\rangle  \hspace{1.8cm} = \,  s'|\xi_{\lambda,k'}^{s'}\rangle\,,
\\
\label{28022016-37-man} && (N_\zeta - N_z)|\xi_{\lambda,k'}^{s'}\rangle \hspace{0.3cm} =\, \lambda |\xi_{\lambda,k'}^{s'}\rangle\,,
\\
\label{28022016-38-man} && (N_{\alpha^\ominussm}  - N_{\alpha^\oplussm})|\xi_{\lambda,k'}^{s'}\rangle =\,  k'|\xi_{\lambda,k'}^{s'}\rangle \,.
\eeq

From relation \rf{28022016-34-man}, we learn that the ket-vector $\xik$ is a
degree-$(s-1)$ homogeneous polynomial in the oscillators $\alpha^a$,
$\alpha^z$, $\zeta$.  Relation \rf{28022016-36-man} tells us that the basis ket-vector $|\xi_{\lambda,k'}^{s'}\rangle$ is a degree-$s'$  homogeneous polynomial in the oscillators $\alpha^a$. From relations \rf{28022016-37-man} and \rf{28022016-38-man} we learn that the ket-vector $|\xi_{\lambda,k'}^{s'}\rangle$ is an eigenvector of the respective operators $N_\zeta - N_z$ and $N_{\alpha^\ominussm}  - N_{\alpha^\oplussm}$.
Also we note that, in terms of the ket-vector $\xik$, constraint given in \rf{28022016-30a1-man} takes the following form
\be \label{28022016-39-man}
\bar{\alpha}^2 \xik = 0 \,.
\ee

\noindent {\bf Remark on gauge symmetries of short conformal fields}. Gauge transformation parameters \rf{28022016-27a-man}, \rf{28022016-31-man} entering the ordinary-derivative formulation of the short conformal fields  have been found in Ref.\cite{Metsaev:2007rw}. In order to match result in Ref.\cite{Metsaev:2007rw} with the presentation in this paper we now write down an explicit form of the ket-vector $\xik$ for gauge transformation parameters of the short conformal field. Such explicit form is obtained by plugging $\kappa=s+\frac{d-4}{2}$ and labels $s'$, $\lambda$, $k'$ given in \rf{28022016-31-man} into ket-vector $\xik$ \rf{28022016-32-man}, \rf{28022016-33-man}. Doing so, we get

\beq
\label{29022016-03-man} && \hspace{-1cm} \xik  =  \sum_{s',k'}
|\xi_{s'+1-s,k'}^{s'}\rangle \,, \hspace{1.8cm}  \hbox{ for short conformal fields},
\\
\label{29022016-03a1-man} && \hspace{-1cm}  |\xi_{s'+1-s,k'}^{s'}\rangle   \equiv
\frac{ \alpha_z^{ s-1-s' } (\alpha^\oplussm)^{\half(k_{s'}+1-k')} (\alpha^\ominussm)^{\half(k_{s'}+1+k')} }{ \sqrt{( s-1-s' )!}\,\, (\frac{k_{s'}+1+k'}{2})!\, s'!}
\alpha^{a_1}\ldots \alpha^{a_{s'}}\xi_{s'+1-s,k'}^{a_1\ldots a_{s'}} |0\rangle\,,\qquad
\eeq
where $k_{s'}$ is given in \rf{28022016-31-man}.

Comparing the basis ket-vectors in \rf{28022016-33-man} and \rf{29022016-03a1-man}, we see that the basis ket-vectors for gauge transformation parameters \rf{29022016-03a1-man}, which are related to gauge symmetries of the short conformal field, do not depend on the oscillator $\zeta$.%
\footnote{ We note the clash of the notation and conventions in this paper and in Ref.\cite{Metsaev:2007rw}. Namely the gauge transformation parameters $\xi_{s'+1-s,k'}^{a_1\ldots a_{s'}}$ appearing in \rf{29022016-03a1-man} are denoted by $\xi_{k'-1}^{a_1\ldots a_{s'}}$  in Ref.\cite{Metsaev:2007rw}. Also we note that the ket-vector $\xik$ in Ref.\cite{Metsaev:2007rw} is obtained from \rf{29022016-03-man}, \rf{29022016-03a1-man} by using the following replacements: $\alpha^z \rightarrow \zeta$, $\alpha^\oplussm \rightarrow \upsilon^\ominussm$, $\alpha^\ominussm \rightarrow \upsilon^\oplussm$. }
This implies that ket-vector $\xik$ \rf{29022016-03-man} satisfies the following algebraic constraints %
\beq
&& (N_\alpha + N_z - s + 1)\xik =0 \,,
\\
&& (N_z  + N_{\alpha^\oplussm}+ N_{\alpha^\ominussm} - k_s)\xik =0 \,,\qquad k_s \equiv s + \frac{d-6}{2}\,,
\eeq
while basis ket-vectors $|\xi_{\lambda,k'}^{s'}\rangle$ \rf{29022016-03a1-man} satisfy the algebraic constraints
\beq
&& N_\alpha |\xi_{s'+1-s,k'}^{s'}\rangle  \hspace{2cm} = \,  s'|\xi_{s'+1-s,k'}^{s'}\rangle\,,
\\
&& N_z|\xi_{s'+1-s,k'}^{s'}\rangle \hspace{1.8cm} =\, (s-1-s') |\xi_{s'+1-s,k'}^{s'}\rangle\,,
\\
&& (N_{\alpha^\ominussm}  - N_{\alpha^\oplussm})|\xi_{s'+1-s,k'}^{s'}\rangle \,\, =\,\,  k'|\xi_{s'+1-s,k'}^{s'}\rangle \,.
\eeq

\noindent {\bf Gauge transformations of conformal fields}. The use of the ket-vector $\phik$ for the description of the conformal fields and the ket-vector $\xik$ for the description of the gauge transformation parameters allows us to present gauge transformations of all conformal fields on an equal footing. Namely, gauge transformations of the long, partial-short, short and special conformal fields can entirely be presented in terms of the ket-vectors $\phik$, $\xik$ in the following way:

\be \label{29022016-03b1-man}
\delta
\phik
= G \xik  \,,
\qquad
G \equiv
\alpar
- e_1
- \alpha^2 \frac{1}{2N_\alpha +d-2}\eb_1\,,
\ee
where the operators  $e_1$, $\eb_1$ appearing in \rf{29022016-03b1-man} are given by relations \rf{28022016-23-man}-\rf{28022016-26-man}.

From \rf{29022016-03b1-man}, we see the following two characteristic features of the gauge transformations in our approach to all conformal fields listed in the Table.

\noindent \ibf) The gauge transformations of fields do not involve higher than first order terms in derivatives.

\noindent \iibf) The one-derivative contributions to the gauge transformations of fields take the form of standard gradient gauge transformations.

%%%%%%%%%%%%%%%%%%%%%%%%%%%%%%%%%%%%%%%%%%%%%%%%%%%%%%%%%%%%%%%%%%%%%%%%%
%%%%%%%%%%%%%%%%%%%%%%%%%%%%%%%%%%%%%%%%%%%%%%%%%%%%%%%%%%%%%%%%%%%%%%%%%
\newsection{\large Realization of conformal symmetries in ordinary-derivative approach}\label{sec-06}
%%%%%%%%%%%%%%%%%%%%%%%%%%%%%%%%%%%%%%%%%%%%%%%%%%%%%%%%%%%%%%%%%%%%%%%%%
%%%%%%%%%%%%%%%%%%%%%%%%%%%%%%%%%%%%%%%%%%%%%%%%%%%%%%%%%%%%%%%%%%%%%%%%%

The dynamics of conformal fields propagating in $R^{d-1,1}$ should respect the conformal algebra $so(d,2)$ symmetries. Note however that, in our ordinary-derivative approach to conformal fields, only the  Lorentz algebra $so(d-1,1)$ symmetries are realized manifestly. This implies that in order to complete our ordinary-derivative formulation of the conformal fields we should provide a realization of the conformal algebra
symmetries on a space of conformal fields.
As we have already said, from relations \rf{02032016-01-man}-\rf{02032016-05-man}, we see that all that is required to complete a description of the conformal symmetries is to find a realization of the operators $\Delta$, $M^{ab}$, and $R^a$ on a space of the ket-vector $\phik$. Our ket-vector $\phik$ is built in terms of the oscillators (see relations \rf{26022016-06-man}, \rf{26022016-07-man}).  For such ket-vector, a realization of the spin operators $M^{ab}$ of the Lorentz algebra is well known and is given by the following relation:
\be \label{02032016-08-man}
M^{ab}
= \alpha^a \bar\alpha^b
- \alpha^b \bar\alpha^a \,.
\ee
A realization of the conformal dimension operator $\Delta$
on a space of the $\phik$ can be read from relations \rf{26022016-05a1-man}, \rf{26022016-12-man}, \rf{27022016-04b5-man} and is given by
\be \label{02032016-09-man}
\Delta  =  \frac{d-2}{2} + \Delta'\,,
\qquad
\Delta'
\equiv N_{\alpha^\ominussm}
- N_{\alpha^\oplussm}
\,.
\ee
Realization of the operator $R^a$ on a space of the ket-vector $\phik$ we find is given
by
\beq
\label{02032016-10-man} R^a & = &   r_{0,1} \alphab^a + \Awt^a \rb_{0,1} + r_{1,1} \partial^a\,,
\\
\label{02032016-11-man} && r_{0,1} = -2\zeta e_\zeta \alpha^\ominussm  + 2 \alpha^z e_z \alphab^\ominussm\,,
\\
\label{02032016-12-man} && \rb_{0,1} =  2\alphab^\ominussm \eb_\zeta \zetab  -  2\alpha^\ominussm e_z \alphab^z
\\
\label{02032016-14-man} && r_{1,1} = -2\alpha^\ominussm \alphab^\ominussm\,,
\eeq
where an operator $\Awt^a$ appearing in \rf{02032016-10-man} is defined in \rf{29022016-10-man}, while the operators $e_\zeta$, $e_z$, $\eb_\zeta$ appearing in \rf{02032016-11-man}, \rf{02032016-12-man} are defined in \rf{28022016-23-man}-\rf{28022016-26-man}.

The following remarks are in order.

\noindent \ibf) We verify that the conformal boost operator $K^a$ \rf{02032016-05-man} with the operator $R^a$ given in \rf{02032016-10-man}-\rf{02032016-14-man} satisfies the commutator $[K^a,K^b]=0$.

\medskip
\noindent \iibf) Using the operators $r_{0,1}$, $\rb_{0,1}$ given in \rf{02032016-11-man}, \rf{02032016-12-man} and the operator $\betabf$ given in \rf{28022016-08-man}-\rf{28022016-10-man}, we verify that, on a space of the ket-vector $\phik$, the operators $r_{0,1}$, $\rb_{0,1}$ \rf{02032016-11-man}, \rf{02032016-12-man} satisfy the following hermitian conjugation rule
\be
(\betabf r_{0,1})^\dagger = - \betabf \rb_{0,1}\,.
\ee

\medskip
\noindent \iiibf) Using relations for the operator $R^a$ presented in \rf{02032016-10-man}-\rf{02032016-14-man}, \rf{28022016-23-man}-\rf{28022016-26-man} and \rf{28022016-21-man}, \rf{28022016-22-man} we verify that the operator $R^a$ is indeed acting on space of ket-vector $\phik$ \rf{26022016-06-man} with the values of the labels $s'$, $\lambda$, $k'$ given in \rf{26022016-01-man}-\rf{26022016-05-man}. Note also that, using the explicit expressions for the operators $r_\zeta$, $r_z$ \rf{28022016-21-man},\rf{28022016-22-man} one can check that the field contents of the special, partial-short, and short conformal fields in \rf{26022016-02-man}-\rf{26022016-05-man} can be realized as invariant subspaces in the field content of the long conformal field in \rf{26022016-01-man}. In this respect there is full analogy with massive and massless fields. As is well known a field content of a massless field can be realized as an invariant subspace in a field content of a massive field when a mass parameter tends to zero. In our case, the field contents of the special, partial-short, and short conformal fields in \rf{26022016-02-man}-\rf{26022016-05-man} are realized as invariant subspaces in the field content of the long conformal field in \rf{26022016-01-man} when the parameter $\kappa$ takes the respective values given in the Table.

\medskip
\noindent \ivbf) A complete ordinary-derivative Lagrangian formulation of a conformal field implies finding a Lagrangian, gauge symmetries and the operator $R^a$. We note that gauge symmetries taken alone do not admit to fix an ordinary-derivative Lagrangian uniquely. It turns out that in order to determine an ordinary-derivative Lagrangian, gauge symmetries, and the operator $R^a$ uniquely we should analyse
restrictions imposed by both gauge symmetries and the conformal algebra $so(d,2)$ symmetries.
The general procedure for finding an ordinary-derivative Lagrangian by using restrictions imposed by gauge symmetries and the conformal algebra $so(d,2)$ symmetries has been developed in Appendix B in Ref.\cite{Metsaev:2007rw}. In the latter reference, we used our general procedure for finding a Lagrangian formulation for the short conformal fields.
Our general procedure in Ref.\cite{Metsaev:2007rw} is applied to the cases of the long, partial-short, and special conformal fields in a rather straightforward way.

\medskip
\noindent \vbf) Our ordinary-derivative approach involves Stueckelberg and auxiliary fields.  The Stueckelberg fields can be removed by using the gauge symmetries in our approach, while the auxiliary fields can be removed by using equations of motion. Doing so, one can make sure that our ordinary-derivative Lagrangian leads to the minimal higher-derivative Lagrangian given in \rf{24022016-05-man}. The general procedure for matching ordinary-derivative and higher-derivative Lagrangian formulations
was developed in Section 5 in Ref.\cite{Metsaev:2007rw}.
Our general procedure in Ref.\cite{Metsaev:2007rw} can straightforwardly be used for the cases of the long, partial-short, and special conformal fields. For the reader's convenience, in Section \ref{sec-07-02} in this paper, by using example of  BRST Lagrangian,  we will demonstrate how higher-derivative BRST Lagrangian is obtained from the ordinary-derivative BRST Lagrangian.

%%%%%%%%%%%%%%%%%%%%%%%%%%%%%%%%%%%%%%%%%%%%%%%%%%%%%%%%%%%%%%%%%%%%%%%
%%%%%%%%%%%%%%%%%%%%%%%%%%%%%%%%%%%%%%%%%%%%%%%%%%%%%%%%%%%%%%%%%%%%%%%
\newsection{ \large BRST Lagrangian and partition functions of conformal fields} \label{sec-07}
%%%%%%%%%%%%%%%%%%%%%%%%%%%%%%%%%%%%%%%%%%%%%%%%%%%%%%%%%%%%%%%%%%%%%%%
%%%%%%%%%%%%%%%%%%%%%%%%%%%%%%%%%%%%%%%%%%%%%%%%%%%%%%%%%%%%%%%%%%%%%%%

In this Section, by using the gauge invariant Lagrangian obtained in Section \ref{sec-04}
and the standard Faddeev-Popov procedure, we obtain a gauge-fixed Lagrangian
of conformal fields which is invariant under global BRST transformations. After this we use the gauge-fixed BRST
Lagrangian for a derivation of partition functions of conformal fields.

%%%%%%%%%%%%%%%%%%%%%%%%%%%%%%%%%%%%%%%%%%%%%%%%%%%%%%%%%%%%%%%%%%%%%%%
%%%%%%%%%%%%%%%%%%%%%%%%%%%%%%%%%%%%%%%%%%%%%%%%%%%%%%%%%%%%%%%%%%%%%%%
\subsection{ BRST Lagrangian of conformal fields}
%%%%%%%%%%%%%%%%%%%%%%%%%%%%%%%%%%%%%%%%%%%%%%%%%%%%%%%%%%%%%%%%%%%%%%%
%%%%%%%%%%%%%%%%%%%%%%%%%%%%%%%%%%%%%%%%%%%%%%%%%%%%%%%%%%%%%%%%%%%%%%%

A general structure of our ordinary-derivative gauge invariant Lagrangian \rf{28022016-02-man} and gauge transformations \rf{29022016-03b1-man} for conformal fields is similar to the one for massive fields.%
\footnote{ Representation of gauge invariant Lagrangian of massive fields in terms of modified de Donder operator was found in Ref.\cite{Metsaev:2008fs}. For review, see Sec.2 in Ref.\cite{Metsaev:2013kaa}.}
For the case of arbitrary spin massive fields, gauge-fixed BRST Lagrangian was obtained in Ref.\cite{Metsaev:2014vda}.
Result in the latter reference is straightforwardly extended to the case of conformal fields.
Let us now to discuss our result for gauge-fixed BRST Lagrangian of conformal fields.

To discuss BRST Lagrangian of conformal fields we introduce a set of Faddeev-Popov fields and Nakanishi-Lautrup fields,
\beq
\label{05032016-01-man} && c_{\lambda,k'}^{a_1\ldots a_{s'}}\,, \qquad \cb_{\lambda,k'}^{a_1\ldots a_{s'}} \hspace{3cm} \hbox{ Faddeev-Popov fields};
\\
&&
\label{05032016-02-man} b_{\lambda,k'}^{a_1\ldots a_{s'}} \hspace{5.2cm}  \hbox{ Nakanishi-Lautrup fields};
\eeq
where, depending on the type of the conformal field, the labels $s',\lambda,k'$ take the same values as the ones for gauge transformation parameters given in \rf{28022016-27-man}-\rf{28022016-31-man}.
In \rf{05032016-01-man}, \rf{05032016-02-man}, the fields with $s'=0$ and $s'=1$ are the respective scalar and vector fields of the Lorentz algebra $so(d-1,1)$, while the fields with $s'>1$ are traceless totally symmetric rank-$s'$ tensor fields of the Lorentz algebra.

To describe the Faddeev-Popov fields and the Nakanishi-Lautrup fields in an easy-to-use form we collect the Faddeev-Popov fields \rf{05032016-01-man} into ket-vectors $\ck$, $\cbk$, while the Nakanishi-Lautrup fields \rf{05032016-02-man} are collected into a ket-vector $\bk$. We note then that the ket-vectors $\ck$, $\cbk$, $\bk$ are obtained by making the respective replacements
\be \label{05032016-03-man}
\xi_{\lambda,k'}^{a_1\ldots a_{s'}} \rightarrow c_{\lambda,k'}^{a_1\ldots a_{s'}}\,, \qquad
\xi_{\lambda,k'}^{a_1\ldots a_{s'}} \rightarrow \cb_{\lambda,k'}^{a_1\ldots a_{s'}}\,, \qquad
\xi_{\lambda,k'}^{a_1\ldots a_{s'}} \rightarrow b_{\lambda,k'}^{a_1\ldots a_{s'}}
\ee
in the expressions for the ket-vector $\xik$ given in \rf{28022016-32-man}, \rf{28022016-33-man}.

Using the ket-vectors above-described, gauge-fixed BRST Lagrangian $\LL_\tot$ can be presented as
\beq
\label{05032016-04-man} && \LL_\tot = \LL + \LL_\qu \,, \qquad  \LL_\qu
= - \langle b
|\Lb\phik
+  \cbbr
( \Box - M^2\bigr)
\ck +
\half \xi
\bbr\bk\,,
\eeq
where the gauge invariant Lagrangian $\LL$ is given in \rf{28022016-02-man}, while the operator $M^2$ and the modified de Donder operator $\Lb$ are given in \rf{28022016-03-man} and \rf{28022016-04-man} respectively.

We now note that gauge-fixed Lagrangian $\LL_\tot$ \rf{05032016-04-man} is invariant not only under BRST but also under  anti-BRST transformations. Namely, gauge-fixed Lagrangian $\LL_\tot$ \rf{05032016-04-man} is invariant under BRST and anti-BRST transformations which take the following form:
\beq
\label{05032016-06-man} && \ssf  \phik
=   G \ck\,,  \qquad
\ssf  \ck  =  0  \,,
\hspace{1.6cm} \ssf  \cbk
=    \bk \,,
\qquad \ssf
\bk = 0 \,,\qquad
\\
\label{05032016-07-man} && \ssfb   \phik
=   G \cbk\,,\qquad
\ssfb   \ck  =
-  \bk \,,\hspace{1cm}
\ssfb   \cbk
= 0\,,
\hspace{1.2cm} \ssfb
\bk  =  0  \,,
\eeq
where an operator $G$ appearing in \rf{05032016-06-man}, \rf{05032016-07-man} is defined in \rf{29022016-03b1-man}. Using \rf{05032016-06-man}, \rf{05032016-07-man}, we verify that the BRST and anti-BRST transformations \rf{05032016-06-man}, \rf{05032016-07-man} are off-shell nilpotent:
\be \label{05032016-08-man}
\ssf^2=0\,,  \qquad \ssfb^2=0\,, \qquad \ssf\ssfb+\ssfb\ssf=0\,.
\ee

For a computation of partition functions of the conformal fields, it is convenient to use the $\xi=1$ gauge.
Doing so, and integrating out the Nakanishi-Lautrup fields, we find that the BRST Lagrangian $\LL_\tot$ \rf{05032016-04-man} takes the following form:
\be \label{05032016-09-man}
\LL_\tot
=    \half  \langle
\phi|(1-\frac{1}{4}\alpha^2\alphab^2)
(\Box - M^2)
|\phi\rangle
+ \cbbr
(\Box - M^2) \ck,
\ee
where the operator $M^2$ is given in \rf{28022016-03-man}. Gauge-fixed Lagrangian  \rf{05032016-09-man} is also  invariant under BRST and anti-BRST transformations which take the following form
\beq
\label{05032016-10-man}  && \ssf
\phik
=
G  \ck\,, \qquad
\ssf  \ck = 0\,,
\hspace{1.7cm}  \ssf
\cbk
=  \Lb
\phik \,,
\\
\label{05032016-11-man} && \ssfb
\phik  =
G  \cbk\,,
\qquad
\ssfb
\ck =
-
\Lb \phik\,,
\qquad \ssfb   \cbk   = 0 \,,
\eeq
where the operators $\Lb$ and $G$ appearing in \rf{05032016-10-man}, \rf{05032016-11-man} are defined in \rf{28022016-04-man} and \rf{29022016-03b1-man} respectively. BRST and anti-BRST transformations given \rf{05032016-10-man}, \rf{05032016-11-man} are also nilpotent \rf{05032016-08-man}. However, in contrast to the transformations given in \rf{05032016-06-man}, \rf{05032016-07-man}, transformations \rf{05032016-10-man}, \rf{05032016-11-man} are nilpotent  only for on-shell Faddeev-Popov fields.

Gauge-fixed BRST Lagrangian \rf{05032016-09-man} can be represented in terms of traceless fields which sometimes turn out to be more convenient for computations. To this end we use the well known decomposition of the double-traceless ket-vector $\phik$ into two traceless ket-vectors $|\phi_{_\I} \rangle$, $|\phi_{_\II} \rangle$,
\beq
\label{16032016-03-man} && \phik = |\phi_{_\I} \rangle   + \alpha^2 \NN |\phi_{_\II} \rangle\,, \qquad  \NN\equiv ((2N_\alpha+d)(2N_\alpha+d-2))^{-1/2}\,,
\\
\label{16032016-04-man} && \bar\alpha^2 |\phi_{_\I} \rangle =0\,,  \qquad \bar\alpha^2 |\phi_{_\II} \rangle =0\,.
\eeq
Plugging decomposition \rf{16032016-03-man} into BRST Lagrangian \rf{05032016-09-man}, we get
\be \label{16032016-05a0-man}
\LL_\tot =
\half \langle \phi_{_\I}
| (\Box - M^2)
|\phi_{_\I} \rangle
- \half \langle \phi_{_\II}
| (\Box-M^2)
|\phi_{_\II} \rangle
+  \langle \cb |
(\Box-M^2)
|c\rangle\,.
\ee

To conclude this Section, we note that it is the use of the modified de Donder operator $\Lb$ in gauge-fixed Lagrangian \rf{05032016-04-man} that allows us to get the simple representations for gauge-fixed BRST Lagrangian given in \rf{05032016-09-man}, \rf{16032016-05a0-man} .

%%%%%%%%%%%%%%%%%%%%%%%%%%%%%%%%%%%%%%%%%%%%%%%%%%%%%%%%%%%%%%%%%%%%%%%
%%%%%%%%%%%%%%%%%%%%%%%%%%%%%%%%%%%%%%%%%%%%%%%%%%%%%%%%%%%%%%%%%%%%%%%
\subsection{ Partition functions and number of D.o.F for conformal fields} \label{sec-07-02}
%%%%%%%%%%%%%%%%%%%%%%%%%%%%%%%%%%%%%%%%%%%%%%%%%%%%%%%%%%%%%%%%%%%%%%%
%%%%%%%%%%%%%%%%%%%%%%%%%%%%%%%%%%%%%%%%%%%%%%%%%%%%%%%%%%%%%%%%%%%%%%%

\noindent {\bf  Partition functions for long and special conformal fields}. For the arbitrary spin-$s$ long, secondary long, and special conformal fields in $R^{d-1,1}$, we find that partition functions take one and same form and are given by
\beq
\label{05032016-14-man}  && Z = \frac{ 1 }{(D^s)^\kappa}\,,
\\
\label{05032016-14a1-man} && D^{n} \equiv \sqrt{\det(-\Box)}\,,
\eeq
where, in \rf{05032016-14a1-man} and below, a quantity $D^{n}$ stands for a determinant of the Laplace operator evaluated on space of rank-$n$ traceless tensor field.
From \rf{05032016-14-man}, we see that numbers of propagating D.o.F for the long, secondary long, and special conformal fields take one and same form and are given by
\be  \label{05032016-15-man}
\nbf^\DoF
= \kappa n_s^{so(d)}\,,
\qquad n_s^{{so(d)}}
\equiv (2s+d-2)
\frac{(s+d-3)!}{(d-2)!s!}\,,
\ee
where $n_s^{so(d)}$ given in \rf{05032016-15-man} is nothing but the dimension of the totally symmetric spin-$s$ irrep of the $so(d)$ algebra. It is well known, that the $n_s^{so(d)}$ describes a number of D.o.F for a spin-$s$ totally symmetric massive field propagating in $(d+1)$ dimensional space-time. In other words, we come to the conclusion that {\it numbers of D.o.F for the spin-$s$ totally symmetric long, secondary long, and special conformal fields that propagate in $d$-dimensional space-time are equal to $\kappa$ times the number of  D.o.F for massive spin-$s$ totally symmetric field that propagates in $(d+1)$ dimensional space-time}. For the case of the long conformal field, the same conclusion was achieved by using light-cone gauge formulation of the long conformal field in Ref.\cite{Metsaev:2015rda}.

\medskip
\noindent {\bf  Partition function for partial-short conformal fields}. For the arbitrary spin-$s$ partial-short conformal field in $R^{d-1,1}$ with arbitrary $d$, we find the following partition function
\be \label{05032016-16-man}
Z = \frac{(D^{s-1-t})^{s+\frac{d-2}{2}} }{(D^s)^\kappa}\,, \qquad \kappa = s + \frac{d-4}{2} - t\,,
\ee
where values of the $\kappa$ for the types I and II partial-short conformal fields are given in the Table.

Using \rf{05032016-16-man}, we find that a number of propagating D.o.F for the partial-short conformal field is given by the relation

\be \label{05032016-17-man}
\nbf^\DoF = \frac{ (2s+d-2)(2s+d-4-2t)(s+d-4)! }{ 2(d-2)!s! } \Bigl( s+d-3 -   s \frac{(s-t)_t}{(s-t+d-3)_t}\Bigr)\,,
\ee
where $(p)_q$ stands for the Pochhammer symbol $\Gamma(p+q)/\Gamma(p)$. We note that the number of D.o.F given in \rf{05032016-17-man} is found by using the relation
\be  \label{27032016-01-man}
\nbf^\DoF = \kappa n_s^{so(d)} - (s+\frac{d-2}{2}) n_{s-1-t}^{so(d)}\,,
\ee
where $n_s^{so(d)}$ is defined in \rf{05032016-15-man}.

For type II partial-short conformal field in $R^{3,1}$ with $\kappa=1$ (maximal-depth partial-short conformal field), partition function \rf{05032016-16-man} and number of D.o.F \rf{05032016-17-man} take the form
\beq
\label{16032016-01-man} && \hspace{-1cm} Z  = \frac{(D^{0})^{s+1} }{(D^s) }\,,  \qquad \qquad  \nbf^\DoF = s(s+1)\,,
\nonumber\\[-12pt]
&& \hspace{3cm} \hbox{ for maximal-depth partial-short conformal field in $R^{3,1}$}\,.
\eeq
The partition function and number of D.o.F given in \rf{16032016-01-man} were first obtained in Ref.\cite{Beccaria:2015vaa}. Thus we see that, for the particular case of $d=4$, $t=s-1$, our result for $Z$ and $\nbf^\DoF$ in \rf{05032016-16-man}, \rf{05032016-17-man} agrees with the result reported in the earlier literature and gives the expressions $Z$ and $\nbf^\DoF$ for arbitrary values of $d$ and $t=1,\ldots, s-1$.

\medskip
\noindent {\bf  Partition function for short conformal field}. Partition function for the arbitrary spin-$s$ short conformal field in $R^{d-,1}$ with arbitrary $d$ is well known,
\be \label{16032016-07-man}
Z =
\frac{(D^{s-1})^{\nu_s + 1}}{(D^s)^{\nu_s}}\,,
\qquad \qquad
\nu_s \equiv s + \frac{d-4}{2}\,.
\ee
For $d=4$ and $d\geq 4$, the partition function \rf{16032016-07-man} was obtained in the respective Ref.\cite{Fradkin:1985am} and \cite{Tseytlin:2013jya}. Derivation of partition function \rf{16032016-07-man} by using the gauge-fixed BRST Lagrangian of the short conformal field may be found in Ref.\cite{Metsaev:2014vda}.
As a side remark, we note that partition function \rf{16032016-07-man} can also be obtained by equating $t=0$ in the partition function of the partial-short conformal field \rf{05032016-16-man}.

Using \rf{16032016-07-man}, we find that  the number of propagating D.o.F for the spin-$s$ short conformal field in $R^{d-1,1}$ is given by the well known relation

\be \label{27032016-02-man}
\nbf^\DoF = \half (d-3)(2s+d-2)(2s+d-4) \frac{ (s+d-4)! }{  (d-2)!s! }\,.
\ee
We note that the number of D.o.F given in \rf{27032016-02-man} is found by using relation
\rf{27032016-01-man}, where we set $t=0$ and use $n_s^{so(d)}$ defined in \rf{05032016-15-man}.

For the case of the short conformal field in $R^{3,1}$, partition function \rf{16032016-07-man} and number of D.o.F \rf{27032016-02-man} take the form
\be
\label{27032016-03-man}  Z = \frac{(D^{s-1})^{s+1} }{(D^s)^s }\,,  \qquad \qquad  \nbf^\DoF = s(s+1)\,,
\qquad  \hbox{ for short conformal field in $R^{3,1}$}.\qquad
\ee

For $d=4$ and $d\geq 4$, the numbers of D.o.F \rf{27032016-03-man} and \rf{27032016-02-man} were found first in the respective Ref.\cite{Fradkin:1985am} and Ref.\cite{Metsaev:2007rw}. In Ref.\cite{Metsaev:2007rw}, the $\nbf^\DoF$ \rf{27032016-02-man} was obtained by counting D.o.F that enter  the ordinary-derivative approach of the short conformal field. For $d\geq 4$, the computation of $\nbf^\DoF$  by using a partition function may be found in Ref.\cite{Tseytlin:2013jya}. As a side remark, we note that, according to \rf{27032016-02-man}, there are no local D.o.F  for short conformal fields in $R^{2,1}$.
Interesting recent discussion of conformal fields in $R^{2,1}$ may be found in Ref.\cite{Linander:2016brv}.

\medskip
\noindent {\bf Comparison of D.o.F for short conformal field and maximal-depth partial-short conformal field in $R^{3,1}$}.
In Ref.\cite{Beccaria:2015vaa}, it was noticed that, in $R^{3,1}$,  the numbers of D.o.F for the short conformal field and the maximal-depth partial-short conformal field coincide (see relations \rf{16032016-01-man}, \rf{27032016-03-man}). Here we would like to demonstrate how our ordinary-derivative approach provides a simple and transparent explanation for this interesting fact. To this end, using a
shortcut $\phi_{k'}^{s'}$ for the $so(3,1)$ algebra vector and tensor fields $\phi_{s'-s,k'}^{a_1\ldots
a_{s'}}$, we note that, for the arbitrary spin-$s$ short conformal field in $R^{3,1}$, the field content appearing in
\rf{26022016-05-man} can be presented as follows:
{\small
$$ \hbox{\sf Field content for spin-$s$ short conformal field in $R^{3,1}$} $$
\be \label{26032016-01-man}
\begin{array}{ccccccccc}
\phi_{1-s}^s &
& \phi_{3-s}^s &
& \ldots &
& \phi_{s-3}^s &
&
\phi_{s-1}^s
\\[12pt]
& \phi_{2-s}^{s-1} &
& \phi_{4-s}^{s-1}
& \ldots
& \phi_{s-4}^{s-1}
&
& \phi_{s-2}^{s-1}
&
\\[12pt]
&
& \ldots
&
& \ldots
&
& \ldots
&
&
\\[12pt]
&
&
& \phi_{-1}^2  &
& \phi_1^2
&
&
&
\\[12pt]
&
&
&
& \phi_0^1
&
&
&
&
\end{array}
\ee }

Accordingly, using a
shortcut $\phi_{k'}^{s'}$ for the $so(3,1)$ algebra vector and tensor fields $\phi_{s-s',k'}^{a_1\ldots
a_{s'}}$, we note that, for the arbitrary spin-$s$ maximal-depth partial-short conformal field in $R^{3,1}$, the field content in \rf{26022016-04-man} can be presented as the follows:
{\small
$$ \hbox{\sf Field content for spin-$s$ maximal-depth partial-short conformal field in $R^{3,1}$} $$
\be  \label{26032016-02-man}
\begin{array}{ccccccccc}
&
&
&
& \phi_0^s&
&
&
&
\\[12pt]
&
&
& \phi_{-1}^{s-1}
&
& \phi_1^{s-1}
&
&
&
\\[12pt]
&
& \ldots
&
& \ldots
&
& \ldots
&
&
\\[12pt]
& \phi_{2-s}^2
&
& \phi_{4-s}^2
& \ldots
& \phi_{s-4}^2
&
&
\phi_{s-2}^2 &
\\[12pt]
\phi_{1-s}^1
&
& \phi_{3-s}^1
&
& \ldots
&
& \phi_{s-3}^1
&
&
\phi_{s-1}^1
\end{array}
\ee
}
From \rf{26032016-01-man}, \rf{26032016-02-man}, we see that, in $R^{3,1}$, scalar fields do not appear in the field contents of the short conformal field and the maximal-depth partial-short conformal field.

Using a light-cone gauge formulation, we verify then that light-cone gauge field contents of the short and maximal-depth partial-short conformal fields take the same respective forms as in \rf{26032016-01-man} and \rf{26032016-02-man}, where all vector and totally symmetric tensor fields of the Lorentz algebra $so(3,1)$ should be replaced by the respective vector and totally symmetric traceless tensor fields of the $so(2)$ algebra. Now taking into account that, for arbitrary $s\geq 1$, a dimension of the totally symmetric spin-$s$ irrep of the $so(2)$ algebra is equal to 2, we see that total numbers of D.o.F for light-cone gauge fields in \rf{26032016-01-man} and \rf{26032016-02-man} coincide and equal to $s(s+1)$.

\medskip
\noindent {\bf Higher-derivative Lagrangian}. For the reader's convenience, we now explain how the partition functions and the numbers of D.o.F above-discussed can be obtained by using gauge-fixed BRST Lagrangian. To this end it is convenient to exclude auxiliary fields and cast the ordinary-derivative BRST Lagrangian \rf{05032016-09-man} into a higher-derivative form. We now explain some details of the derivation of a higher-derivative BRST Lagrangian.

First, we note that Lagrangian \rf{05032016-09-man}  leads to the following equations of motion for the basis ket-vectors \rf{26022016-07-man}:
\beq
\label{16032016-08-man}  && \Box |\phi_{\lambda, k'}^{s'}\rangle = |\phi_{\lambda, k'+2}^{s'}\rangle \,, \qquad k' \in  [1-\nu_\lambda,\nu_\lambda -3]_2\,, \qquad  \nu_\lambda \equiv  \kappa + \lambda\,.
\eeq
Equations \rf{16032016-08-man} tell us that all basis ket-vectors \rf{26022016-07-man} having $k' > 1-\nu_\lambda$ can be expressed in terms of the basis ket-vector $|\phi_{\lambda, 1-\nu_\lambda}^{s'}\rangle$ as
\be \label{16032016-08a2-man}
|\phi_{\lambda, k'}^{s'}\rangle  = \Box^{\half(\nu_\lambda - 1 + k')} |\phi_{\lambda, 1- \nu_\lambda}^{s'}\rangle \,,\qquad k' \in  [3-\nu_\lambda,\nu_\lambda -1]_2\,.
\ee
Repeating the same analysis for the Faddeev-Popov fields entering Lagrangian \rf{05032016-09-man}, we find the following solution for auxiliary Faddeev-Popov fields
\be \label{16032016-09-man}
|c_{\lambda, k'}^{s'}\rangle  = \Box^{\half(\nu_\lambda - 1 + k')} |c_{\lambda, 1-\nu_\lambda}^{s'}\rangle  \,, \qquad
|\cb_{\lambda, k'}^{s'}\rangle  = \Box^{\half(\nu_\lambda - 1 + k')} |\cb_{\lambda, 1-\nu_\lambda}^{s'}\rangle  \,,\qquad  k' \in  [3-\nu_\lambda,\nu_\lambda -1]_2\,.
\ee
Thus, from \rf{16032016-08a2-man}, \rf{16032016-09-man}, we see that all gauge fields and Faddeev-Popov fields that  have $k' > 1-\nu_\lambda$ can be expressed in terms of the respective fields $|\phi_{\lambda, 1-\nu_\lambda}^{s'}\rangle$ and $|c_{\lambda, 1-\nu_\lambda}^{s'}\rangle$, $|\cb_{\lambda, 1-\nu_\lambda}^{s'}\rangle$. In other words, all basis ket-vectors that have $k' > 1- \nu_\lambda$ turn out to be auxiliary fields. Taking this into account, we introduce ket-vectors defined by the following  relations:
\beq
\label{16032016-10-man} && |\phi_{_\Lrm}\rangle  \equiv  \sum_{s', \lambda}  |\phi_{\lambda,1-\nu_\lambda}^{s'}\rangle\,,
\\[-3pt]
\label{16032016-11-man} && |c_{_\Lrm}\rangle  \equiv  \sum_{s', \lambda}  |c_{\lambda,1-\nu_\lambda}^{s'}\rangle\,, \qquad |\cb_{_\Lrm}\rangle  \equiv  \sum_{s', \lambda}  |\cb_{\lambda,1-\nu_\lambda}^{s'}\rangle\,,
\eeq
where the ket-vectors $|\phi_{\lambda,1-\nu_\lambda}^{s'}\rangle$ in \rf{16032016-10-man} are defined as in \rf{26022016-07-man}, while the ket-vectors $|c_{\lambda,1-\nu_\lambda}^{s'}\rangle$,
$|\cb_{\lambda,1-\nu_\lambda}^{s'}\rangle$ in \rf{16032016-11-man} are defined as in \rf{28022016-33-man}.
We note that the summation indices $s'$, $\lambda$ in \rf{16032016-10-man} take the values given in \rf{26022016-01-man}-\rf{26022016-05-man}, while
the summation indices $s'$, $\lambda$ in \rf{16032016-11-man} take the values given in \rf{28022016-27-man}-\rf{28022016-31-man}.
Plugging the solution for auxiliary fields \rf{16032016-08a2-man}, \rf{16032016-09-man} into \rf{05032016-09-man}, we find that gauge-fixed BRST Lagrangian \rf{05032016-09-man} can be expressed in terms of ket-vectors \rf{16032016-10-man}, \rf{16032016-11-man} as
\be \label{16032016-14-man}
\LL_\tot
=    \half  \langle
\phi_{_\Lrm}|(1-\frac{1}{4}\alpha^2\alphab^2)
\Box^\nu
|\phi_{_\Lrm}\rangle
+ \langle \cb_{_\Lrm}|
\Box^\nu | c_{_\Lrm}\rangle \,, \qquad \nu \equiv \kappa + N_\zeta - N_z \,.
\ee
Obviously, Lagrangian \rf{16032016-14-man} involves higher derivatives. Thus we see that, after excluding the auxiliary gauge fields and the auxiliary Faddeev-Popov fields, our ordinary-derivative Lagrangian \rf{05032016-09-man} leads to higher-derivative Lagrangian \rf{16032016-14-man}.

Lagrangian \rf{16032016-14-man} can be cast into the form which is more convenient for a computation of partition functions. To this end we use the decomposition of the double-traceless ket-vector $|\phi_{_\Lrm}\rangle$ into two traceless ket-vectors $|\phi_{_{\Lrm\I}} \rangle$, $|\phi_{_{\Lrm\II}} \rangle$ as in \rf{16032016-03-man},
\be \label{16032016-15-man}
|\phi_{_\Lrm}\rangle  = |\phi_{_{\Lrm\I}} \rangle   + \alpha^2 \NN |\phi_{_{\Lrm\II}} \rangle\,, \qquad
\bar\alpha^2 |\phi_{_{\Lrm\I}} \rangle =0\,,  \qquad \bar\alpha^2 |\phi_{_{\Lrm\II}} \rangle =0\,.
\ee
Plugging the decomposition \rf{16032016-15-man} into Lagrangian \rf{16032016-14-man}, we get
\be \label{16032016-05-man}
\LL_\tot =
\half \langle \phi_{_{\Lrm\I}}
| \Box^\nu
|\phi_{_{\Lrm\I}} \rangle
- \half \langle \phi_{_{\Lrm\II}}
| \Box^\nu
|\phi_{_{\Lrm\II}} \rangle
+    \langle \cb_{_\Lrm} |
\Box^\nu
|c_{\Lrm}\rangle\,.
\ee
For the illustration purposes, we note that Lagrangian \rf{16032016-05-man} can be expressed in terms of the scalar, vector, and traceless tensor fields as
\beq
\label{17032016-01-man} &&   \LL_\tot  = \sum_{s',\lambda } \LL_{\I,\lambda}^{s'} - \sum_{s',\lambda \atop s\rightarrow s-2, d\rightarrow d+ 4} \LL_{\II,\lambda}^{s'} +
\sum_{s',\lambda \atop s\rightarrow s-1, d\rightarrow d+ 2} \LL_{\FP,\lambda}^{s'}\,, \qquad
\\
\label{17032016-02-man} &&     \LL_{\I,\lambda}^{s'}  \equiv \frac{1}{2s'!}\, \phi_{\I\lambda,1-\nu_\lambda}^{a_1 \ldots a_{s'}}\beta_{s',\lambda}^\Irm \Box^{\nu_\lambda}  \phi_{\I\lambda,1-\nu_\lambda}^{a_1 \ldots a_{s'}}\,,
\qquad   \LL_{\II,\lambda}^{s'}  \equiv  \frac{1}{2s'!}\, \phi_{\II\lambda,1-\nu_\lambda}^{a_1 \ldots a_{s'}} \beta_{s',\lambda}^\IIrm \Box^{\nu_\lambda}    \phi_{\II\lambda,1-\nu_\lambda}^{a_1 \ldots a_{s'}}\,,\qquad
\\
\label{17032016-03-man} &&   \LL_{\FP,\lambda}^{s'}  \equiv \frac{1}{s'!}\, \cb_{\lambda,1-\nu_\lambda}^{a_1 \ldots a_{s'}} \beta_{s',\lambda}^{\rm FP} \Box^{\nu_\lambda}   c_{\lambda,1-\nu_\lambda}^{a_1 \ldots a_{s'}}\,, \qquad \nu_\lambda \equiv \kappa + \lambda\,.
\eeq
The summation indices $s'$, $\lambda$ for the $\LL_{\I,\lambda}^{s'}$-terms in \rf{17032016-01-man} take values given in \rf{26022016-01-man}-\rf{26022016-05-man}, while the summation indices $s'$, $\lambda$ for the $\LL_{\II,\lambda}^{s'}$-terms and the $\LL_{\FP,\lambda}^{s'}$-terms in \rf{17032016-01-man} take values which are obtained by making the respective replacements $s\rightarrow s-2$, $d\rightarrow d+ 4$ and $s\rightarrow s-1$, $d\rightarrow d+ 2$ in \rf{26022016-01-man}-\rf{26022016-05-man}.  Quantities $\beta_{s',\lambda}^\Irm$, $\beta_{s',\lambda}^\IIrm$, and $\beta_{s',\lambda}^{\rm FP}$ appearing in \rf{17032016-02-man}, \rf{17032016-03-man} are obtained by using the respective substitutions $N_\zeta \rightarrow \half (s-s'+\lambda)$,  $N_\zeta \rightarrow  \half (s-2-s'+\lambda)$, and $N_\zeta \rightarrow \half (s-1-s'+\lambda)$ in expressions for $\betabf$ given in \rf{28022016-08-man}-\rf{28022016-10-man}.

Using Lagrangian \rf{17032016-01-man} allows us to obtain a general representation for a partition function.
Namely, we see that Lagrangian \rf{17032016-01-man} leads to the following representation for a partition function
\beq
\label{17032016-04-man} && Z = \frac{Z_{s-1,d+2}Z_{s-1,d+2}}{Z_{s,d}Z_{s-2,d+4}}\,,
\\
\label{17032016-05-man} && Z_{s,d} \equiv  \prod_{s',\lambda } (D^{s'})^{\kappa+\lambda}\,,
\eeq
where, depending on the type of the conformal field,  the product indices $s'$, $\lambda$ appearing in \rf{17032016-05-man} take values given in \rf{26022016-01-man}-\rf{26022016-05-man}.

We now demonstrate how the general expressions given in \rf{17032016-04-man}, \rf{17032016-05-man} can be used for the computation of partition functions of various conformal fields.

\noindent {\bf Computation of $Z$ for long conformal field}. For the case of the long conformal field, the labels $s'$, $\lambda$ take values shown in \rf{26022016-01-man}. This implies that $Z_{s,d}$ \rf{17032016-05-man} takes the form
\be
\label{17032016-06-man}  Z_{s,d} \equiv  \prod_{s'\in [0,s]_1} \,\,\, \prod_{\scriptstyle \lambda\in [s-s']_2 } (D^{s'})^{\kappa+\lambda}\,.
\ee
where the relations $k\in [p,q]_1$, $k\in [p]_2$ are defined in \rf{29022016-05a1-man}-\rf{29022016-05a3-man}. Using the general formula
\beq
\label{17032016-07-man} \prod_{\scriptstyle\lambda\in [p,q]_2 }   (D^{s'})^{\kappa+\lambda}
= (D^{s'})^{ (\frac{q-p}{2} + 1)(\kappa+ \frac{p+q}{2}) }\,,
\eeq
where the notation $\lambda\in [p,q]_2$ is defined in \rf{29022016-05a3-man}, we see that $Z_{s,d}$ \rf{17032016-06-man} can be represented as
\be \label{17032016-08-man}
Z_{s,d} = \prod_{s'\in [0,s]_1} (D^{s'})^{ (s-s' + 1)\kappa }\,.
\ee
Plugging \rf{17032016-08-man} in \rf{17032016-04-man}, we obtain $Z$ given in \rf{05032016-14-man}. Note that for the long conformal field the $Z_{s,d}$ \rf{17032016-08-man} does not depend on $d$ explicitly. Repeating the  above-described computation for the case of the secondary long and special conformal fields, we get the same $Z$ as in \rf{05032016-14-man}.

\noindent {\bf Computation of partition functions for partial-short conformal fields}. Partition functions for the type I and type II partial-short conformal fields take one and same form given in \rf{05032016-16-man}. Computation of a partition function turns out to be more involved for the case of the type II partial-short conformal fields in $R^{d-1,1}$, $d\geq 6$. Therefore, for the reader's convenience, we consider this case. To this end we note that, for the type II partial-short conformal field, one has the relation $t > (d-6)/2$, where $t$ is defined in \rf{05032016-16-man}. We verify then that the domain of values of the labels $s'$, $\lambda$ given in \rf{26022016-04-man} can be represented as a direct sum of three domains of values of labels denoted by (1), (2), and (3) and defined by the following relations:
\beq
\label{17032016-09-man} &&(1): \qquad s'\in [0,t-\frac{d-6}{2}]_1\,, \hspace{1.7cm}  \lambda\in [2t-s-s'-d+6,s'-s+2t]_2\,, \qquad
\\
\label{17032016-10-man} &&(2): \qquad s'\in [t-\frac{d-8}{2},s-t]_1 \,, \hspace{1cm} \lambda\in [s'-s,s'-s+2t]_2\,,
\\
\label{17032016-11-man} &&(3):\qquad  s'\in [s-t+1,s]_1\,, \hspace{1.8cm}  \lambda\in [s-s']_2 \,,
\eeq
where the relations $k\in [p]_2$, $k\in [p,q]_1$, and $k\in [p,q]_2$ are defined in \rf{29022016-05a1-man}-\rf{29022016-05a3-man}, while the parameter $t$ is defined in \rf{05032016-16-man}.
Accordingly, the general expression for the partition function $Z$ given in \rf{17032016-04-man}, \rf{17032016-05-man} can be represented as
\beq
 \label{17032016-14-man} && Z = Z^{(1)} Z^{(2)} Z^{(3)}\,,
\\
\label{17032016-15-man} && Z^{(k)} \equiv \frac{ Z_{s-1,d+2}^{(k)}Z_{s-1,d+2}^{(k)} }{ Z_{s,d}^{(k)} Z_{s-2,d+4}^{(k)} }\,, \qquad k =1,2,3\,,
\\
\label{17032016-16-man} && Z_{s,d}^{(1)} \equiv \prod_{s'\in [0,t-\frac{d-6}{2}]_1}\,\,\, \prod_{\scriptstyle\lambda\in [2t-s-s'-d+6,s'-s+2t]_2 }   (D^{s'})^{\kappa+\lambda}\,,
\\
\label{17032016-17-man} && Z_{s,d}^{(2)} \equiv \prod_{s'\in [t-\frac{d-8}{2},s-t]_1}\,\,\, \prod_{\scriptstyle\lambda\in [s'-s,s'-s+2t]_2 }   (D^{s'})^{\kappa+\lambda}\,,
\\
\label{17032016-18-man} && Z_{s,d}^{(3)} \equiv \prod_{s'\in [s-t+1,s]_1}\,\,\, \prod_{\scriptstyle\lambda\in [s-s']_2 }   (D^{s'})^{\kappa+\lambda} \,.
\eeq
Using general formula \rf{17032016-07-man}, we straightforwardly compute the products over indices $\lambda$ appearing in \rf{17032016-16-man}-\rf{17032016-18-man},
\beq
\label{17032016-19-man} && \hspace{-1.3cm} \prod_{\scriptstyle\lambda\in [2t-s-s'-d+6,s'-s+2t]_2 }   (D^{s'})^{\kappa+\lambda}
= (D^{s'})^{ (t + 1)(s'+ \frac{d-4}{2}) }\,,
\\
\label{17032016-20-man} && \hspace{-0.6cm} \prod_{\scriptstyle\lambda\in [s'-s,s'-s+2t]_2 }   (D^{s'})^{\kappa+\lambda}
= (D^{s'})^{ (t + 1)(s'+ \frac{d-4}{2}) }\,,
\\
\label{17032016-21-man} && \prod_{\scriptstyle\lambda\in [s-s']_2 }   (D^{s'})^{\kappa+\lambda} = (D^{s'})^{ (s-s' + 1) \kappa }\,.
\eeq
Plugging \rf{17032016-19-man}-\rf{17032016-21-man} into \rf{17032016-16-man}-\rf{17032016-18-man}, we get the relations
\be
\label{17032016-22-man}  Z_{s,d}^{(1)} Z_{s,d}^{(2)}  =   \prod_{\scriptstyle s'\in [0,s-t]_1 }   (D^{s'})^{ (t + 1)(s'+ \frac{d-4}{2}) }\,, \qquad\quad
 Z_{s,d}^{(3)} =\prod_{s'\in [s-t+1,s]_1}\,\,\,  (D^{s'})^{ (s-s' + 1) \kappa }\,.\qquad
\ee
Using the definition of $Z^{(k)}$ given in  \rf{17032016-15-man} and relations \rf{17032016-22-man}, we find
\be \label{17032016-24-man}
Z^{(1)} Z^{(2)} = \frac{ (D^{s-1-t})^{(t+1)(\kappa+1)} }{ (D^{s-t})^{(t+1)\kappa }  } \,, \hspace{2.4cm}
Z^{(3)} = \frac{ (D^{s-t})^{(t+1)\kappa} }{ (D^s)^\kappa (D^{s-1-t})^{ t \kappa} } \,.\qquad\qquad
\ee
Relations \rf{17032016-14-man} and \rf{17032016-24-man} lead to the partition function for the partial-short conformal fields given in \rf{05032016-16-man}. For the derivation of \rf{05032016-16-man}, the interrelation between $\kappa$ and $t$ given in \rf{05032016-16-man} also should be used.

%%%%%%%%%%%%%%%%%%%%%%%%%%%%%%%%%%%%%%%%%%%%%%%%%%%%%%%%%%%%%%%%%%%%%%%
%%%%%%%%%%%%%%%%%%%%%%%%%%%%%%%%%%%%%%%%%%%%%%%%%%%%%%%%%%%%%%%%%%%%%%%
\newsection{ \large Conclusions}\label{sec-08}
%%%%%%%%%%%%%%%%%%%%%%%%%%%%%%%%%%%%%%%%%%%%%%%%%%%%%%%%%%%%%%%%%%%%%%%
%%%%%%%%%%%%%%%%%%%%%%%%%%%%%%%%%%%%%%%%%%%%%%%%%%%%%%%%%%%%%%%%%%%%%%%

In the framework of AdS/CFT correspondence, conformal field that propagates in $R^{d-1,1}$ and has conformal dimension as in \rf{24022016-01-man} is dual to a non-normalizable mode of bulk field that propagates in $AdS_{d+1}$ and has lowest eigenvalue of an energy operator equal to $E_0 = \kappa + \frac{d}{2}$.%
\footnote{ For arbitrary values of $\kappa$, $s$, and $d$, action for massive spin-$s$ AdS field evaluated on a solution of the Dirichlet problem was first found in Refs.\cite{Metsaev:2011uy,Metsaev:2010zu}. Interesting discussion of group theoretical aspects of AdS/CFT correspondence may be found in Refs.\cite{Dobrev:1998md}.}
This implies that the short, partial-short and long conformal fields in $R^{d-1,1}$ are dual to the respective massless, partial-massless and massive fields in $AdS_{d+1}$.
Taking this into account, we speculated \cite{Metsaev:2014sfa} on some special regime in $AdS$ superstring theory  when   parameters $\kappa$ for all massive higher-spin fields take integer values.
One can expect that such conjectured regime in the AdS superstring theory should be related via AdS/CFT correspondence to stringy theory of conformal fields that involves low-spin short conformal fields and higher-spin long conformal fields.%
\footnote{ One can speculate on the similar regime in the higher-spin gauge theory \cite{Vasiliev:1990en} considered in the perspective of broken higher-spin symmetries. The recent interesting discussion of broken higher-spin symmetries may be found in Ref.\cite{Skvortsov:2015pea} (see also Refs.\cite{Giombi:2016hkj})}
In fact it is such conjectured regime in AdS superstring theory that triggered our interest to the study of long conformal fields.

In this paper, we developed the ordinary-derivative Lagrangian formulation for all totally symmetric conformal fields propagating in $R^{d-1,1}$. Though, at the present time, the minimal Lagrangian formulation of conformal fields, which with exception of some particular cases involves higher derivatives, is more popular, we think that the ordinary-derivative approach is more perspective. This is to say that in our ordinary-derivative approach the gauge symmetries of conformal fields are realized, among other things, by using Stueckelberg fields. Stueckelberg fields
turned out be useful for the study of string theory. Namely, all Lorentz covariant formulations of
string theory available in the literature have been built by exploiting Stueckelberg fields. Stueckelberg fields turn also to be helpful for study of field theoretical models of interacting massive AdS fields (see, e.g., Refs.\cite{Zinoviev:2010av,Metsaev:2006ui}). We think therefore that the use of gauge symmetries involving Stueckelberg fields might also be useful for the study of various problems of conformal fields. The use of gauge symmetries involving Stueckelberg fields for building ordinary-derivative Lagrangian of the interacting spin-2 conformal field in six dimensions may be found in Ref.\cite{Metsaev:2010kp}. In this respect it would be interesting to extend result in the latter reference to higher space-time dimensions along the lines in Refs.\cite{Boulanger:2007ab}.

We now discuss various potentially interesting
generalizations of our approach and review some related studies in the literature.

\noindent \ibf) In Ref.\cite{Metsaev:2015yyv}, we developed ordinary-derivative
BRST-BV approach to the totally symmetric short conformal fields. Generalization of result in Ref.\cite{Metsaev:2015yyv} to the case of the long, partial-short and special conformal fields should be relatively straightforward.
We note also BRST--BV higher-derivative Lagrangian for the Weyl gravity (spin-2
conformal field) was discussed in Ref.\cite{Boulanger:2001he}, while
a higher-derivative gauge-fixed BRST Lagrangian of bosonic arbitrary spin
short conformal field was found in Ref.\cite{Metsaev:2014vda}.
Application of BRST--BV approach for studying equations of motion for bosonic arbitrary spin
conformal fields may be found in Ref.\cite{Bekaert:2013zya}.
We note also that ordinary-derivative formulation of conformal fields and gauge invariant formulation  of massive fields have many common features. BRST approach has been extensively used for studying massless and massive fields (see, e.g., Refs.\cite{Buchbinder:2005ua}). We think that a use of the methods in Refs.\cite{Buchbinder:2005ua} should lead to better understanding of BRST formulation of conformal fields.

\noindent \iibf) An extended hamiltonian formulation of field dynamics is one of powerful approaches in modern field theory. The extended hamiltonian formulation of the arbitrary spin bosonic short-conformal fields was developed in Ref.\cite{Metsaev:2011iz} by using ordinary-derivative gauge invariant Lagran\-gian. Our ordinary-derivative gauge invariant Lagrangian for the long, partial-short and special conformal fields is similar to the one for the short-conformal fields. We expect therefore that generalization of result in Ref.\cite{Metsaev:2011iz} to the case of the long, partial-short and special conformal fields should be straightforward.
Recent interesting discussion of extended hamiltonian formulation of conformal fields may be found in Ref.\cite{Henneaux:2015cda}.

\noindent \iiibf) Mixed-symmetry fields enter a spectrum of the string theory field content.
Obviously that the conjectured stringy generalization of higher-spin conformal field theory should also involve mixed-symmetry conformal fields. In the frame-like approach, mixed-symmetry conformal fields were studied in Ref.\cite{Vasiliev:2009ck}. Also, in the latter reference, the minimal Lagrangian for arbitrary mixed-symmetry long, short, partial-short, and special conformal fields has been found. It would be interesting to apply approach
in Ref.\cite{Vasiliev:2009ck} to the study of ordinary-derivative Lagrangian of mixed-symmetry conformal fields. In the framework of light-cone gauge approach, Lagrangian formulation of mixed-symmetry long and short conformal field in $R^{3,1}$ may be found in Ref.\cite{Metsaev:2014sfa} (see also Ref.\cite{Metsaev:2015rda}). In the framework of ambient approach, mixed-symmetry short conformal fields were studied in Ref.\cite{Chekmenev:2015kzf}. Ordinary-derivative formulation of self-dual conformal fields may be found in Ref.\cite{Metsaev:2008ba}

\noindent \ivbf) In this paper, we have discussed a Lagrangian formulation of conformal fields by using the double-traceless fields. In the literature, a formulation of the dynamics of massless gauge fields in terms of unconstrained triplets was discussed in Refs.\cite{Bengtsson:1986ys} (see also recent interesting discussion in Ref.\cite{Agugliaro:2016ngl}). A use of  unconstrained triplets might lead to simple Lagrangian formulation of conformal fields.

\noindent \vbf) In this paper we considered conformal fields in flat space. It would be interesting to extend
our approach for the long conformal fields to the case of AdS space \cite{Metsaev:2014iwa} and consider various applications of the ordinary-derivative formulation along the lines in Refs.\cite{Florakis:2014aaa,Ananth:2012tf}.

\bigskip
{\bf Acknowledgments}.  This work was supported by the Russian Science Foundation grant 14-42-00047.


\small
\begin{thebibliography}{50}

\parskip=-3pt



%\cite{Fradkin:1985am}
\bibitem{Fradkin:1985am}
  E.~S.~Fradkin and A.~A.~Tseytlin,
  %``Conformal Supergravity,''
  Phys.\ Rept.\  {\bf 119}, 233 (1985).
  %%CITATION = PRPLC,119,233;%%



%\cite{Segal:2002gd}
\bibitem{Segal:2002gd}
  A.~Y.~Segal,
  %``Conformal higher spin theory,''
  Nucl.\ Phys.\ B {\bf 664}, 59 (2003)
  [arXiv:hep-th/0207212].
  %%CITATION = HEP-TH 0207212;%%


%\cite{Vasiliev:2009ck}
\bibitem{Vasiliev:2009ck}
  M.~A.~Vasiliev,
  %``Bosonic conformal higher-spin fields of any symmetry,''
  Nucl.\ Phys.\  B {\bf 829}, 176 (2010)
  [arXiv:0909.5226 [hep-th]].
  %%CITATION = NUPHA,B829,176;%%




%\cite{Marnelius:2009uw}
\bibitem{Marnelius:2009uw}
  R.~Marnelius,
  ``Lagrangian higher spin field theories from the O(N) extended supersymmetric particle,''
  arXiv:0906.2084 [hep-th].
  %%CITATION = ARXIV:0906.2084;%%



%\cite{Liu:1998ty}
\bibitem{Liu:1998ty}
  H.~Liu and A.~A.~Tseytlin,
  %``On four-point functions in the CFT/AdS correspondence,''
  Phys.\ Rev.\  D {\bf 59}, 086002 (1999)
  [arXiv:hep-th/9807097].
  %%CITATION = PHRVA,D59,086002;%%


%\cite{Metsaev:2009ym}
\bibitem{Metsaev:2009ym}
  R.~R.~Metsaev,
  %``Gauge invariant two-point vertices of shadow fields, AdS/CFT, and conformal
  %fields,''
  Phys.\ Rev.\  D {\bf 81}, 106002 (2010)
  [arXiv:0907.4678 [hep-th]].
  %%CITATION = PHRVA,D81,106002;%%


%\cite{Metsaev:2015rda}
\bibitem{Metsaev:2015rda}
  R.~R.~Metsaev,
  %``Light-cone AdS/CFT-adapted approach to AdS fields/currents, shadows, and conformal fields,''
  JHEP {\bf 1510}, 110 (2015)
%  doi:10.1007/JHEP10(2015)110
  [arXiv:1507.06584 [hep-th]].
  %%CITATION = doi:10.1007/JHEP10(2015)110;%%


%\cite{Iorio:1996ad}
\bibitem{Iorio:1996ad}
  A.Iorio, L.O'Raifeartaigh, I.Sachs, C.Wiesendanger,
  %``Weyl gauging and conformal invariance,''
  Nucl.Phys.B {\bf 495}, 433 (1997)
% doi:10.1016/S0550-3213(97)00190-9
  [hep-th/9607110].
  %%CITATION = doi:10.1016/S0550-3213(97)00190-9;%%


%\cite{Erdmenger:1997wy}
\bibitem{Erdmenger:1997wy}
  J.~Erdmenger and H.~Osborn,
  %``Conformally covariant differential operators: Symmetric tensor fields,''
  Class.\ Quant.\ Grav.\  {\bf 15}, 273 (1998)
%  doi:10.1088/0264-9381/15/2/003
  [gr-qc/9708040].
  %%CITATION = doi:10.1088/0264-9381/15/2/003;%%



%\cite{Metsaev:2007fq}
\bibitem{Metsaev:2007fq}
  R.~R.~Metsaev,
  %``Ordinary-derivative formulation of conformal low-spin fields,''
  JHEP {\bf 1201}, 064 (2012)
  [arXiv:0707.4437 [hep-th]].
  %%CITATION = JHEPA,1201,064;%%

%\cite{Metsaev:2007rw}
\bibitem{Metsaev:2007rw}
  R.~R.~Metsaev,
  %``Ordinary-derivative formulation of conformal totally symmetric arbitrary spin bosonic fields,''
  JHEP {\bf 1206}, 062 (2012)
% doi:10.1007/JHEP06(2012)062
  [arXiv:0709.4392 [hep-th]].
  %%CITATION = doi:10.1007/JHEP06(2012)062;%%





%\cite{Metsaev:2008fs}
\bibitem{Metsaev:2008fs}
  R.~R.~Metsaev,
  %``Shadows, currents and AdS,''
  Phys.\ Rev.\ D {\bf 78}, 106010 (2008)
%  doi:10.1103/PhysRevD.78.106010
  [arXiv:0805.3472 [hep-th]].
  %%CITATION = doi:10.1103/PhysRevD.78.106010;%%




%\cite{Barut:1982nj}
\bibitem{Barut:1982nj}
  A.~O.~Barut and B.~W.~Xu,
  %``On Conformally Covariant Spin-2 And Spin 3/2 Equations,''
  J.\ Phys.\ A {\bf 15}, L207 (1982).
% doi:10.1088/0305-4470/15/4/010
  %%CITATION = doi:10.1088/0305-4470/15/4/010;%%


%\cite{Drew:1980yk}
\bibitem{Drew:1980yk}
  M.~S.~Drew and J.~D.~Gegenberg,
  %``Conformally Covariant Massless Spin-2 Field Equations,''
  Nuovo Cim.\ A {\bf 60}, 41 (1980).
% doi:10.1007/BF02776555
  %%CITATION = doi:10.1007/BF02776555;%%



%\cite{Deser:1983tm}
\bibitem{Deser:1983tm}
  S.~Deser and R.~I.~Nepomechie,
  %``Anomalous Propagation of Gauge Fields in Conformally Flat Spaces,''
  Phys.\ Lett.\ B {\bf 132}, 321 (1983).
%  doi:10.1016/0370-2693(83)90317-9
  %%CITATION = doi:10.1016/0370-2693(83)90317-9;%%



%\cite{Bekaert:2013zya}
\bibitem{Bekaert:2013zya}
  X.~Bekaert and M.~Grigoriev,
  %``Higher order singletons, partially massless fields and their boundary values in the ambient approach,''
  Nucl.\ Phys.\ B {\bf 876}, 667 (2013)  [arXiv:1305.0162 [hep-th]].
  %%CITATION = ARXIV:1305.0162;%%
%
\\
%
%\cite{Bekaert:2012vt}
%\bibitem{Bekaert:2012vt}
  X.~Bekaert and M.~Grigoriev,
  %``Notes on the ambient approach to boundary values of AdS gauge fields,''
  J.\ Phys.\ A {\bf 46}, 214008 (2013)
  [arXiv:1207.3439 [hep-th]].
  %%CITATION = ARXIV:1207.3439;%%




%\cite{Barnich:2015tma}
\bibitem{Barnich:2015tma}
  G.Barnich, X.Bekaert, M.Grigoriev,
  %``Notes on conformal invariance of gauge fields,''
  J. Phys. A {\bf 48}, no.50, 505402 (2015)
%  doi:10.1088/1751-8113/48/50/505402
  [arXiv:1506.00595 [hep-th]].
  %%CITATION = doi:10.1088/1751-8113/48/50/505402;%%



%\cite{Metsaev:2011uy}
\bibitem{Metsaev:2011uy}
  R.~R.~Metsaev,
  %``Anomalous conformal currents, shadow fields and massive AdS fields,''
  Phys.\ Rev.\ D {\bf 85}, 126011 (2012)
%  doi:10.1103/PhysRevD.85.126011
  [arXiv:1110.3749 [hep-th]].
  %%CITATION = doi:10.1103/PhysRevD.85.126011;%%




%\cite{Freedman:1998tz}
\bibitem{Freedman:1998tz}
  D.Z.Freedman, S.D.Mathur, A.Matusis and L.Rastelli,
  %``Correlation functions in the CFT($d$)/AdS($d+1$) correspondence,''
  Nucl.Phys. B {\bf 546}, 96 (1999)
  [hep-th/9804058].
  %%CITATION = NUPHA,B546,96;%%
%
\\
%
%\cite{Mueck:1998iz}
%\bibitem{Mueck:1998iz}
  W.~Mueck and K.~S.~Viswanathan,
  %``Conformal field theory correlators from classical field theory on  anti-de
  %Sitter space. II: Vector and spinor fields,''
  Phys.\ Rev.\  D {\bf 58}, 106006 (1998)
  [arXiv:hep-th/9805145].
  %%CITATION = PHRVA,D58,106006;%%
%
\\
%
%\cite{Polishchuk:1999nh}
%\bibitem{Polishchuk:1999nh}
  A.~Polishchuk,
  %``Massive symmetric tensor field on AdS,''
  JHEP {\bf 9907}, 007 (1999)
  hep-th/9905048.
  %%CITATION = JHEPA,9907,007;%%



%\cite{Aref'eva:2014mia}
\bibitem{Aref'eva:2014mia}
  I.~Y.~Aref'eva,
  %``Holographic approach to quark-gluon plasma in heavy ion collisions,''
  Phys.\ Usp.\  {\bf 57}, 527 (2014).
%  doi:10.3367/UFNe.0184.201406a.0569
  %%CITATION = doi:10.3367/UFNe.0184.201406a.0569;%%
%
\\
%
%\cite{Aref'eva:1998nn}
%\bibitem{Aref'eva:1998nn}
  I.~Y.~Aref'eva and I.~V.~Volovich,
  ``On large N conformal theories, field theories in anti-de Sitter space  and
  singletons,''
  arXiv:hep-th/9803028.
  %%CITATION = HEP-TH/9803028;%%





%\cite{Alkalaev:2012ic}
\bibitem{Alkalaev:2012ic}
  K.~Alkalaev,
  %``Massless hook field in AdS(d+1) from the holographic perspective,''
  JHEP {\bf 1301}, 018 (2013)
%  doi:10.1007/JHEP01(2013)018
  [arXiv:1210.0217 [hep-th]].
  %%CITATION = doi:10.1007/JHEP01(2013)018;%%



%\cite{Deser:2001xr}
\bibitem{Deser:2001xr}
  S.~Deser and A.~Waldron,
  % ``Null propagation of partially massless higher spins in (A)dS and
  %cosmological constant speculations,''
  Phys.\ Lett.\ B {\bf 513}, 137 (2001)
  [arXiv:hep-th/0105181].
  %%CITATION = HEP-TH 0105181;%%
%
\\
%
%\cite{Deser:2001us}
%\bibitem{Deser:2001us}
  S.~Deser and A.~Waldron,
  %``Partial masslessness of higher spins in (A)dS,''
  Nucl.\ Phys.\ B {\bf 607}, 577 (2001)
  [arXiv:hep-th/0103198].
  %%CITATION = HEP-TH 0103198;%%


%\cite{Zinoviev:2001dt}
\bibitem{Zinoviev:2001dt}
  Y.~M.~Zinoviev,
  ``On massive high spin particles in AdS,''
  hep-th/0108192.
  %%CITATION = HEP-TH/0108192;%%


  %\cite{Metsaev:2006zy}
\bibitem{Metsaev:2006zy}
  R.~R.~Metsaev,
  %``Gauge invariant formulation of massive totally symmetric fermionic fields in (A)dS space,''
  Phys.\ Lett.\ B {\bf 643}, 205 (2006)
%  doi:10.1016/j.physletb.2006.11.002
  [hep-th/0609029].
  %%CITATION = doi:10.1016/j.physletb.2006.11.002;%%


%\cite{Skvortsov:2006at}
\bibitem{Skvortsov:2006at}
  E.~D.~Skvortsov and M.~A.~Vasiliev,
  %``Geometric formulation for partially massless fields,''
  Nucl.\ Phys.\ B {\bf 756}, 117 (2006)
%  doi:10.1016/j.nuclphysb.2006.06.019
  [hep-th/0601095].
  %%CITATION = doi:10.1016/j.nuclphysb.2006.06.019;%%
%
\\
%
%\cite{Skvortsov:2009nv}
%\bibitem{Skvortsov:2009nv}
  E.~D.~Skvortsov,
  %``Gauge fields in (A)dS within the unfolded approach: algebraic aspects,''
  JHEP {\bf 1001}, 106 (2010)
  [arXiv:0910.3334 [hep-th]].
  %%CITATION = JHEPA,1001,106;%%
%
\\
%
%\cite{Alkalaev:2009vm}
%\bibitem{Alkalaev:2009vm}
  K.~B.~Alkalaev and M.~Grigoriev,
  %``Unified BRST description of AdS gauge fields,''
  Nucl.\ Phys.\ B {\bf 835}, 197 (2010)
% doi:10.1016/j.nuclphysb.2010.04.004
  [arXiv:0910.2690 [hep-th]].
  %%CITATION = doi:10.1016/j.nuclphysb.2010.04.004;%%
%
\\
%
%\cite{Alkalaev:2011zv}
%\bibitem{Alkalaev:2011zv}
  K.~Alkalaev and M.~Grigoriev,
  %``Unified BRST approach to (partially) massless and massive AdS fields of arbitrary symmetry type,''
  Nucl.\ Phys.\ B {\bf 853}, 663 (2011)
%  doi:10.1016/j.nuclphysb.2011.08.005
  [arXiv:1105.6111 [hep-th]].
  %%CITATION = doi:10.1016/j.nuclphysb.2011.08.005;%%



%\cite{Joung:2014aba}
\bibitem{Joung:2014aba}
  E.~Joung, W.~Li and M.~Taronna,
  %``No-Go Theorems for Unitary and Interacting Partially Massless Spin-Two Fields,''
  Phys.\ Rev.\ Lett.\  {\bf 113}, 091101 (2014)
%  doi:10.1103/PhysRevLett.113.091101
  [arXiv:1406.2335 [hep-th]].
  %%CITATION = doi:10.1103/PhysRevLett.113.091101;%%
%
\\
%
%\cite{Joung:2012hz}
%\bibitem{Joung:2012hz}
  E.~Joung, L.~Lopez and M.~Taronna,
  %``Generating functions of (partially-)massless higher-spin cubic interactions,''
  JHEP {\bf 1301}, 168 (2013)
%  doi:10.1007/JHEP01(2013)168
  [arXiv:1211.5912 [hep-th]].
%%CITATION = doi:10.1007/JHEP01(2013)168;%%
%
\\
%
%\cite{Zinoviev:2014zka}
%\bibitem{Zinoviev:2014zka}
  Y.~M.~Zinoviev,
  %``Massive spin-2 in the Fradkin–Vasiliev formalism. I. Partially massless case,''
  Nucl.\ Phys.\ B {\bf 886}, 712 (2014)
%  doi:10.1016/j.nuclphysb.2014.07.013
  [arXiv:1405.4065 [hep-th]].
  %%CITATION = doi:10.1016/j.nuclphysb.2014.07.013;%%




%\cite{Ponomarev:2016jqk}
\bibitem{Ponomarev:2016jqk}
  D.~Ponomarev and A.~A.~Tseytlin,
  %``On quantum corrections in higher-spin theory in flat space,''
  arXiv:1603.06273 [hep-th].
  %%CITATION = ARXIV:1603.06273;%%
%
\\
%
%\cite{Francia:2007qt}
%\bibitem{Francia:2007qt}
  D.~Francia, J.~Mourad and A.~Sagnotti,
  %``Current Exchanges and Unconstrained Higher Spins,''
  Nucl.\ Phys.\ B {\bf 773}, 203 (2007)
%  doi:10.1016/j.nuclphysb.2007.03.021
  [hep-th/0701163].
  %%CITATION = doi:10.1016/j.nuclphysb.2007.03.021;%%
%
\\
%
%\cite{Guttenberg:2008qe}
%\bibitem{Guttenberg:2008qe}
  S.~Guttenberg and G.~Savvidy,
  %``Schwinger-Fronsdal Theory of Abelian Tensor Gauge Fields,''
  SIGMA {\bf 4}, 061 (2008)
  [arXiv:0804.0522 [hep-th]].
  %%CITATION = ARXIV:0804.0522;%%
%
\\
%
%\cite{Manvelyan:2008ks}
%\bibitem{Manvelyan:2008ks}
  R.~Manvelyan, K.~Mkrtchyan and W.~Ruhl,
  %``Ultraviolet behaviour of higher spin gauge field propagators and one loop
  %mass renormalization,''
  Nucl.\ Phys.\  B {\bf 803}, 405 (2008)
  [arXiv:0804.1211 [hep-th]].
  %%CITATION = NUPHA,B803,405;%%
%
\\
%
%\cite{Fotopoulos:2009iw}
%\bibitem{Fotopoulos:2009iw}
  A.~Fotopoulos and M.~Tsulaia,
  %``Current Exchanges for Reducible Higher Spin Multiplets and Gauge Fixing,''
  JHEP {\bf 0910}, 050 (2009)
  [arXiv:0907.4061 [hep-th]].
  %%CITATION = JHEPA,0910,050;%%


%\cite{Joung:2015eny}
\bibitem{Joung:2015eny}
  E.~Joung, S.~Nakach and A.~A.~Tseytlin,
  %``Scalar scattering via conformal higher spin exchange,''
  JHEP {\bf 1602}, 125 (2016)
%  doi:10.1007/JHEP02(2016)125
  [arXiv:1512.08896 [hep-th]].
  %%CITATION = doi:10.1007/JHEP02(2016)125;%%



%\cite{Metsaev:2013kaa}
\bibitem{Metsaev:2013kaa}
  R.~R.~Metsaev,
  %``Light-cone gauge approach to arbitrary spin fields, currents, and shadows,''
  J.\ Phys.\ A {\bf 47}, 375401 (2014)
%  doi:10.1088/1751-8113/47/37/375401
  [arXiv:1312.5679 [hep-th]].
  %%CITATION = doi:10.1088/1751-8113/47/37/375401;%%




%\cite{Beccaria:2015vaa}
\bibitem{Beccaria:2015vaa}
  M.~Beccaria and A.~A.~Tseytlin,
  %``On higher spin partition functions,''
  J.\ Phys.\ A {\bf 48}, no. 27, 275401 (2015)
%  doi:10.1088/1751-8113/48/27/275401
  [arXiv:1503.08143 [hep-th]].
  %%CITATION = doi:10.1088/1751-8113/48/27/275401;%%



%\cite{Tseytlin:2013jya}
\bibitem{Tseytlin:2013jya}
  A.~A.~Tseytlin,
  %``On partition function and Weyl anomaly of conformal higher spin fields,''
  Nucl.\ Phys.\ B {\bf 877}, 598 (2013)
%  doi:10.1016/j.nuclphysb.2013.10.009
  [arXiv:1309.0785 [hep-th]].
  %%CITATION = doi:10.1016/j.nuclphysb.2013.10.009;%%
%
\\
%
%\cite{Tseytlin:2013fca}
%\bibitem{Tseytlin:2013fca}
  A.~A.~Tseytlin,
  %``Weyl anomaly of conformal higher spins on six-sphere,''
  Nucl.\ Phys.\ B {\bf 877}, 632 (2013)
  [arXiv:1310.1795 [hep-th]].
  %%CITATION = ARXIV:1310.1795;%%



%\cite{Metsaev:2014vda}
\bibitem{Metsaev:2014vda}
  R.~R.~Metsaev,
  %``BRST invariant effective action of shadow fields, conformal fields, and AdS/CFT,''
  Theor.\ Math.\ Phys.\  {\bf 181}, no. 3, 1548 (2014)
%  doi:10.1007/s11232-014-0235-1
  [arXiv:1407.2601 [hep-th]].
  %%CITATION = doi:10.1007/s11232-014-0235-1;%%




%\cite{Linander:2016brv}
\bibitem{Linander:2016brv}
  H.~Linander and B.~E.~W.~Nilsson,
  %``The non-linear coupled spin 2 - spin 3 Cotton equation in three dimensions,''
  arXiv:1602.01682 [hep-th].


%\cite{Metsaev:2010zu}
\bibitem{Metsaev:2010zu}
  R.~R.~Metsaev,
  %``Gauge invariant approach to low-spin anomalous conformal currents and
  %shadow fields,''
  Phys.\ Rev.\  D {\bf 83}, 106004 (2011)
  [arXiv:1011.4261 [hep-th]].
  %%CITATION = PHRVA,D83,106004;%%



%\cite{Dobrev:1998md}
\bibitem{Dobrev:1998md}
V.~K.~Dobrev,
%``Intertwining operator realization of the AdS/CFT correspondence,''
Nucl.\ Phys.\ B {\bf 553}, 559 (1999) [arXiv:hep-th/9812194];
%%CITATION = HEP-TH 9812194;%%
%
\\
%
%\cite{Aizawa:2014yqa}
%\bibitem{Aizawa:2014yqa}
  N.~Aizawa and V.~K.~Dobrev,
  %``Intertwining Operator Realization of anti de Sitter Holography,''
  Rept.\ Math.\ Phys.\  {\bf 75}, 179 (2015)
%  doi:10.1016/S0034-4877(15)30002-1
  [arXiv:1406.2129 [hep-th]].
  %%CITATION = doi:10.1016/S0034-4877(15)30002-1;%%


%\cite{Metsaev:2014sfa}
\bibitem{Metsaev:2014sfa}
  R.~R.~Metsaev,
  %``Mixed-symmetry fields in AdS(5), conformal fields, and AdS/CFT,''
  JHEP {\bf 1501}, 077 (2015)
%  doi:10.1007/JHEP01(2015)077
  [arXiv:1410.7314 [hep-th]].
  %%CITATION = doi:10.1007/JHEP01(2015)077;%%


%\cite{Vasiliev:1990en}
\bibitem{Vasiliev:1990en}
  M.~A.~Vasiliev,
  %``Consistent equation for interacting gauge fields of all spins in (3+1)-dimensions,''
  Phys.\ Lett.\ B {\bf 243}, 378 (1990).
%  doi:10.1016/0370-2693(90)91400-6
  %%CITATION = doi:10.1016/0370-2693(90)91400-6;%%
%
\\
%
%\cite{Vasiliev:2003ev}
%\bibitem{Vasiliev:2003ev}
  M.~A.~Vasiliev,
  %``Nonlinear equations for symmetric massless higher spin fields in
  %(A)dS(d),''
  Phys.\ Lett.\  B {\bf 567}, 139 (2003)
  [arXiv:hep-th/0304049].
  %%CITATION = PHLTA,B567,139;%%





%\cite{Skvortsov:2015pea}
\bibitem{Skvortsov:2015pea}
  E.D.Skvortsov,
  On (Un)Broken Higher-Spin Symmetry in Vector Models,
  arXiv:1512.05994 [hep-th].
  %%CITATION = ARXIV:1512.05994;%%


%\cite{Giombi:2016hkj}
\bibitem{Giombi:2016hkj}
  S.~Giombi and V.~Kirilin,
%  ``Anomalous dimensions in CFT with weakly broken higher spin symmetry,''
  arXiv:1601.01310 [hep-th].
  %%CITATION = ARXIV:1601.01310;%%
%
\\
%
%\cite{Hikida:2016wqj}
%\bibitem{Hikida:2016wqj}
  Y.~Hikida,
%  ``The masses of higher spin fields on AdS(4) and conformal perturbation theory,''
  arXiv:1601.01784 [hep-th].


%\cite{Zinoviev:2010av}
\bibitem{Zinoviev:2010av}
  Yu.~M.~Zinoviev,
  %``On electromagnetic interactions for massive mixed symmetry field,''
  JHEP {\bf 1103}, 082 (2011)
  [arXiv:1012.2706 [hep-th]].
  %%CITATION = JHEPA,1103,082;%%
%
\\
%
%\cite{Zinoviev:2006im}
%\bibitem{Zinoviev:2006im}
  Yu.~M.~Zinoviev,
  %``On massive spin 2 interactions,''
  Nucl.\ Phys.\  B {\bf 770}, 83 (2007)
  [arXiv:hep-th/0609170].
  %%CITATION = NUPHA,B770,83;%%


%\cite{Metsaev:2006ui}
\bibitem{Metsaev:2006ui}
  R.~R.~Metsaev,
  %``Gravitational and higher-derivative interactions of massive spin 5/2 field
  %in (A)dS space,''
  Phys.\ Rev.\  D {\bf 77}, 025032 (2008)
  [arXiv:hep-th/0612279].
  %%CITATION = PHRVA,D77,025032;%%



%\cite{Metsaev:2010kp}
\bibitem{Metsaev:2010kp}
R.~R.~Metsaev,
%``6d conformal gravity,''
J.\ Phys.\ A  {\bf 44}, 175402 (2011) [arXiv:1012.2079 [hep-th]].
%%CITATION = JPAGB,A44,175402;%%



%\cite{Boulanger:2007ab}
\bibitem{Boulanger:2007ab}
  N.~Boulanger,
  %``Algebraic Classification of Weyl Anomalies in Arbitrary Dimensions,''
  Phys.\ Rev.\ Lett.\  {\bf 98}, 261302 (2007)
%  doi:10.1103/PhysRevLett.98.261302
  [arXiv:0706.0340 [hep-th]].
  %%CITATION = doi:10.1103/PhysRevLett.98.261302;%%
%
\\
%
%\cite{Boulanger:2004zf}
%\bibitem{Boulanger:2004zf}
  N.~Boulanger and J.~Erdmenger,
  %``A Classification of local Weyl invariants in D=8,''
  Class.\ Quant.\ Grav.\  {\bf 21}, 4305 (2004)
%  doi:10.1088/0264-9381/21/18/003
  [hep-th/0405228].
  %%CITATION = doi:10.1088/0264-9381/21/18/003;%%



%\cite{Metsaev:2015yyv}
\bibitem{Metsaev:2015yyv}
  R.~R.~Metsaev,
% ``BRST-BV approach to conformal fields,''
J.\ Phys.\ A  {\bf 49}, 175401 (2016)
  arXiv:1511.01836 [hep-th].




%\cite{Boulanger:2001he}
\bibitem{Boulanger:2001he}
  N.~Boulanger and M.~Henneaux,
  %``A derivation of Weyl gravity,''
  Annalen Phys.\  {\bf 10}, 935 (2001)
  [arXiv:hep-th/0106065].
  %%CITATION = ANPYA,10,935;%%


%\cite{Buchbinder:2005ua}
\bibitem{Buchbinder:2005ua}
  I.~L.~Buchbinder and V.~A.~Krykhtin,
  %``Gauge invariant Lagrangian construction for massive bosonic higher spin
  %fields in D dimensions,''
  Nucl.\ Phys.\ B {\bf 727}, 537 (2005)
  [arXiv:hep-th/0505092].
  %%CITATION = HEP-TH 0505092;%%
%
\\
%
%\cite{Buchbinder:2007vq}
%\bibitem{Buchbinder:2007vq}
  I.L. Buchbinder, V.Krykhtin and A.Reshetnyak,
  %``BRST approach to Lagrangian construction for fermionic higher spin   fields
  %in AdS space,''
  Nucl.Phys. B {\bf 787}, 211 (2007)
  [hep-th/0703049].
  %%CITATION = NUPHA,B787,211;%%
%
\\
%
%\cite{Buchbinder:2006ge}
%\bibitem{Buchbinder:2006ge}
  I.~L.~Buchbinder, V.~A.~Krykhtin and P.~M.~Lavrov,
  %``\hfill{\normalsize{}hep-th/0608005,''
  Nucl.\ Phys.\  B {\bf 762}, 344 (2007)
  hep-th/0608005
  %%CITATION = NUPHA,B762,344;%%
%
\\
%
%\cite{Buchbinder:2011xw}
%\bibitem{Buchbinder:2011xw}
  I.~L.~Buchbinder and A.~Reshetnyak,
  %``General Lagrangian Formulation for Higher Spin Fields with Arbitrary Index Symmetry. I. Bosonic fields,''
  Nucl.\ Phys.\ B {\bf 862}, 270 (2012)
%  doi:10.1016/j.nuclphysb.2012.04.016
  [arXiv:1110.5044 [hep-th]].
  %%CITATION = doi:10.1016/j.nuclphysb.2012.04.016;%%
%
\\
%
%\cite{Moshin:2007jt}
%\bibitem{Moshin:2007jt}
  P.~Y.~Moshin and A.~A.~Reshetnyak,
  %``BRST approach to Lagrangian formulation for mixed-symmetry fermionic
  %higher-spin fields,''
  JHEP {\bf 0710}, 040 (2007)
  [arXiv:0707.0386 [hep-th]].
  %%CITATION = JHEPA,0710,040;%%
%
\\
%
%\cite{Polyakov:2015usr}
%\bibitem{Polyakov:2015usr}
  D.~Polyakov,
  %``Higher Spins at the Quintic Order: Localization Effect and Simplifications,''
  Phys.\ Rev.\ D {\bf 93}, no. 4, 045001 (2016)
%  doi:10.1103/PhysRevD.93.045001
  [arXiv:1511.04563 [hep-th]].
  %%CITATION = doi:10.1103/PhysRevD.93.045001;%%




%\cite{Metsaev:2011iz}
\bibitem{Metsaev:2011iz}
  R.~R.~Metsaev,
  %``Extended Hamiltonian Action for Arbitrary Spin Fields in Flat And AdS Spaces,''
  J.\ Phys.\ A {\bf 46}, 214021 (2013)
%  doi:10.1088/1751-8113/46/21/214021
  [arXiv:1112.0976 [hep-th]].
  %%CITATION = doi:10.1088/1751-8113/46/21/214021;%%


  %\cite{Henneaux:2015cda}
\bibitem{Henneaux:2015cda}
  M.~Henneaux, S.~Hörtner and A.~Leonard,
  %``Higher Spin Conformal Geometry in Three Dimensions and Prepotentials for Higher Spin Gauge Fields,''
  JHEP {\bf 1601}, 073 (2016)
%  doi:10.1007/JHEP01(2016)073
  [arXiv:1511.07389 [hep-th]].
  %%CITATION = doi:10.1007/JHEP01(2016)073;%%


%\cite{Chekmenev:2015kzf}
\bibitem{Chekmenev:2015kzf}
  A.~Chekmenev and M.~Grigoriev,
%  ``Boundary values of mixed-symmetry massless fields in AdS space,''
  arXiv:1512.06443 [hep-th].



%\cite{Metsaev:2008ba}
\bibitem{Metsaev:2008ba}
  R.~R.~Metsaev,
  %``Conformal self-dual fields,''
  J.\ Phys.\ A  {\bf 43}, 115401 (2010)
  [arXiv:0812.2861 [hep-th]].
  %%CITATION = JPAGB,A43,115401;%%



%\cite{Bengtsson:1986ys}
\bibitem{Bengtsson:1986ys}
  A.~K.~H.~Bengtsson,
  %``A Unified Action for Higher Spin Gauge Bosons From Covariant String Theory,''
  Phys.\ Lett.\ B {\bf 182}, 321 (1986).
  doi:10.1016/0370-2693(86)90100-0
  %%CITATION = doi:10.1016/0370-2693(86)90100-0;%%
%
\\
%
%\cite{Francia:2002pt}
%\bibitem{Francia:2002pt}
  D.~Francia and A.~Sagnotti,
  %``On the geometry of higher spin gauge fields,''
  Class.\ Quant.\ Grav.\  {\bf 20}, S473 (2003)
%  [Comment.\ Phys.\ Math.\ Soc.\ Sci.\ Fenn.\  {\bf 166}, 165 (2004)]
%  [PoS JHW {\bf 2003}, 005 (2003)]
%  doi:10.1088/0264-9381/20/12/313
   [hep-th/0212185].
\\
%
%\cite{Sagnotti:2003qa}
%\bibitem{Sagnotti:2003qa}
  A.~Sagnotti and M.~Tsulaia,
  %``On higher spins and the tensionless limit of string theory,''
  Nucl.\ Phys.\  B {\bf 682}, 83 (2004)
  [arXiv:hep-th/0311257].
  %%CITATION = NUPHA,B682,83;%%


%\cite{Agugliaro:2016ngl}
\bibitem{Agugliaro:2016ngl}
  A.~Agugliaro, F.~Azzurli and D.~Sorokin,
%  ``Fermionic higher-spin triplets in AdS,''
  arXiv:1603.02251 [hep-th].
  %%CITATION = ARXIV:1603.02251;%%



%\cite{Metsaev:2014iwa}
\bibitem{Metsaev:2014iwa}
  R.~R.~Metsaev,
  %``Arbitrary spin conformal fields in (A)dS,''
  Nucl.\ Phys.\ B {\bf 885}, 734 (2014)
%  doi:10.1016/j.nuclphysb.2014.06.013
  [arXiv:1404.3712 [hep-th]].
  %%CITATION = doi:10.1016/j.nuclphysb.2014.06.013;%%



%\cite{Florakis:2014aaa}
\bibitem{Florakis:2014aaa}
  I.~Florakis, D.~Sorokin and M.~Tsulaia,
  %``Higher Spins in Hyper-Superspace,''
  Nucl.\ Phys.\ B {\bf 890}, 279 (2014)
%  doi:10.1016/j.nuclphysb.2014.11.017
  [arXiv:1408.6675 [hep-th]].
  %%CITATION = doi:10.1016/j.nuclphysb.2014.11.017;%%
%
\\
%
%\cite{Florakis:2014kfa}
%\bibitem{Florakis:2014kfa}
  I.~Florakis, D.~Sorokin and M.~Tsulaia,
  %``Higher Spins in Hyperspace,''
  JHEP {\bf 1407}, 105 (2014)
  [arXiv:1401.1645 [hep-th]].
  %%CITATION = ARXIV:1401.1645;%%


%\cite{Ananth:2012tf}
\bibitem{Ananth:2012tf}
  S.~Ananth, S.~Kovacs and S.~Parikh,
  %``Gauge-invariant correlation functions in light-cone superspace,''
  JHEP {\bf 1205}, 096 (2012)
%  doi:10.1007/JHEP05(2012)096
  [arXiv:1203.5376 [hep-th]].
  %%CITATION = doi:10.1007/JHEP05(2012)096;%%
%
\\
%
%\cite{Ghodsi:2014hua}
%\bibitem{Ghodsi:2014hua}
  A.~Ghodsi, B.~Khavari and A.~Naseh,
  %``Holographic Two-Point Functions in Conformal Gravity,''
  JHEP {\bf 1501}, 137 (2015)
%  doi:10.1007/JHEP01(2015)137
  [arXiv:1411.3158 [hep-th]].
  %%CITATION = doi:10.1007/JHEP01(2015)137;%%




\end{thebibliography}
\end{document}